%**start of header
\input amstex
\documentstyle{amsppt}
\magnification=\magstephalf
%%%%%%%%%%%% changes to amsppt.sty %%%%%%%%%%%%%%%%%%%%%%%
 \addto\tenpoint{\baselineskip 15pt
  \abovedisplayskip18pt plus4.5pt minus9pt
  \belowdisplayskip\abovedisplayskip
  \abovedisplayshortskip0pt plus4.5pt
  \belowdisplayshortskip10.5pt plus4.5pt minus6pt}\tenpoint
\pagewidth{6.5truein} \pageheight{8.9truein}
\subheadskip\bigskipamount
\belowheadskip\bigskipamount
\aboveheadskip=3\bigskipamount
\catcode`\@=11
\def\output@{\shipout\vbox{%
 \ifrunheads@ \makeheadline \pagebody
       \else \pagebody \fi \makefootline 
 }%
 \advancepageno \ifnum\outputpenalty>-\@MM\else\dosupereject\fi}
\outer\def\subhead#1\endsubhead{\par\penaltyandskip@{-100}\subheadskip
  \noindent{\subheadfont@\ignorespaces#1\unskip\endgraf}\removelastskip
  \nobreak\medskip\noindent}
\outer\def\enddocument{\par% \par will do a runaway check for \endref
  \add@missing\endRefs
  \add@missing\endroster \add@missing\endproclaim
  \add@missing\enddefinition
  \add@missing\enddemo \add@missing\endremark \add@missing\endexample
 \ifmonograph@ % do nothing
 \else
 \vfill
 \nobreak
 \thetranslator@
 \count@\z@ \loop\ifnum\count@<\addresscount@\advance\count@\@ne
 \csname address\number\count@\endcsname
 \csname email\number\count@\endcsname
 \repeat
\fi
 \supereject\end}
\catcode`\@=\active
%%%%%%%%%%%%%%% other macros %%%%%%%%%%%%%%%%%%%%%%%%%%%%
\CenteredTagsOnSplits
\NoBlackBoxes
\nologo
\def\today{\ifcase\month\or
 January\or February\or March\or April\or May\or June\or
 July\or August\or September\or October\or November\or December\fi
 \space\number\day, \number\year}
\define\({\left(}
\define\){\right)}
\define\Ahat{{\hat A}}

\define\CC{{\Bbb C}}
\define\CP{{\Bbb C\Bbb P}}

\define\HH{{\Bbb H}}
\define\Hom{\operatorname{Hom}}
\define\Map{\operatorname{Map}}

\define\RR{{\Bbb R}}

\define\Spin{\operatorname{Spin}}

\define\ZZ{{\Bbb Z}}
\define\[{\left[}
\define\]{\right]}
\define\ch{\operatorname{ch}}
\define\chiup{\raise.5ex\hbox{$\chi$}}
\define\cir{S^1}

%\define\exertag #1#2{\removelastskip\bigskip\medskip\eightpoint\noindent%
%\hbox{\rm\ignorespaces#2\unskip} #1.\ }  
\define\exertag #1#2{#2\ #1}

\define\inv{^{-1}}
\define\mstrut{^{\vphantom{1*\prime y}}}
\define\protag#1 #2{#2\ #1}
\define\rank{\operatorname{rank}}
\define\res#1{\negmedspace\bigm|_{#1}}
\define\temsquare{\raise3.5pt\hbox{\boxed{ }}}

\define\theprotag#1 #2{#2~#1}

\define\xca#1{\removelastskip\medskip\noindent{\smc%
#1\unskip.}\enspace\ignorespaces }

\define\zmod#1{\ZZ/#1\ZZ}

\define\zt{\zmod2}

\NoRunningHeads % USE IN FINAL VERSION; THEN COMMENT OUT NEXT LINE
% \headline{\eightpoint PRELIMINARY VERSION \hfil \today}

\define\AhG{\Ahat_{\Gamma }}
\define\Aht{\check{\Cal{A}}}
\define\Ah{\check{A}}
\define\BhG{\check{B}_\Gamma }
\define\Bh{\check{B}}
\define\ChG{\check{C}_\Gamma }
\define\Ch{\check{C}}
\define\Det{\operatorname{Det}}
\define\Eh{\check E}
\define\Euler{\operatorname{Euler}}
\define\Fh{\check{F}}
\define\Ghb{\Gh^{\bul}}
\define\Ghq#1{\check{\Gamma}(#1)}
\define\Gh{\check{\Gamma}}
\define\Hh{\check{H}}
\define\KOh{{{KO}\spcheck}}
\define\KSph{{{KSp}\spcheck}}
\define\Khat{\hat{K}}
\define\Kh{\check{K}}
\define\MSB{M\Spin\wedge BO\langle8\rangle}
\define\PP{\Bbb{P}}
\define\Qb{\overline{Q}}
\define\RZ{\RR/\ZZ}
\define\RetN{\RR_{@,@,@,@,\tau} \times N}
\define\RtN{\RR_{@,@,@,@,@,t}\times N}
\define\Sym{\operatorname{Sym}}
\define\TTh{\check{\Cal{T}}}

\define\TT{\Bbb{T}}
\define\Th{\check T}
\define\ZhG{\check{Z}_\Gamma}

\define\Zh{\check{Z}}

\define\agb{A_{\Gamma}^{\bul}}
\define\ah{\check{a}}
\define\alh{\check{\alpha }}
\define\at{\tilde{a}}
\define\bh{\check{b}}
\define\bul{\bullet}

\define\chh{\check{c}}

\define\clf{\Omega _{\text{cl}}}
\define\curv{\operatorname{curv}}
\define\ddt{\tilde{\dd}}
\define\dd{\bold{d}}
\define\eh{\check{\epsilon }}
\define\fb{\bar{f}}
\define\filt{\operatorname{filtration}}
\define\gbb{\overline{\Gamma}^{\bul}}
\define\gcb{\gcoh\bul}
\define\gcoh#1{\Gamma ^{#1}}
\define\gh{\check{\gamma }}
\define\jh{\check{\jmath}}
\define\kh{\check{k}}
\define\kinetic{\operatorname{kinetic}}
\define\lh{\check{\lambda }}
\define\mh{\check{\mu }}
\define\pbring{\pi _{-\bul}\Gamma _\RR}
\define\pfaff{\operatorname{pfaff}}
\define\ph{\hat{\rho }}
\define\pring{\pi \Gamma _\RR}
\define\psh{\psi_{1/2}}
\define\qhb{\bar{\qh}}
\define\qh{\check{q}}
\define\ring{\gcb(pt)}
\define\rring{\Gamma _{\RR}}
\define\scrL{\Cal{L}}
\define\spinc{$\text{spin}^c$}
\define\sqo{\sqrt{-1}}
\define\th{\check{\tau }}
\define\tp{2\pi }
\define\univ{_{\text{univ}}}
\define\wb{\check{w}}
\define\wh{\check{w}}
\define\zh{\check{\zeta }}

\refstyle{A}
\widestnumber\key{SSSSS}   % for widest bibliography name
\document
%**end of header

 \pretitle{$$\boxed{\boxed{\text{REVISED VERSION}}}$$\par\vskip 3pc}

	\topmatter
 \title\nofrills Dirac Charge Quantization and Generalized Differential
Cohomology \endtitle 
 \author Daniel S. Freed  \endauthor
 \thanks The author is supported by NSF grant DMS-0072675.\endthanks
 \affil Department of Mathematics \\ University of Texas at Austin\endaffil 
 \address Department of Mathematics, University of Texas, Austin, TX
78712\endaddress 
 %\curraddr \endcurraddr
 \email dafr\@math.utexas.edu \endemail
 \date February 5, 2001\enddate
 \dedicatory To the Gang Of Four\enddedicatory
 %\abstract \endabstract
	\endtopmatter

\document

The {\it classical\/} Maxwell equations, which describe electricity and
magnetism in four-dimensional spacetime, may be generalized in many
directions.  For example, nonabelian generalizations play an important role
in both geometry and physics.  The equations also admit abelian
generalizations in which differential forms of degree greater than two come
into play.  Such forms enter into high dimensional supergravity theories, so
also into string theory and M-theory.  There are analogs of electric and
magnetic currents for these higher degree forms.  In the classical theory
these are also differential forms, and their de Rham cohomology classes in
real cohomology (with support conditions) are the corresponding electric and
magnetic charges.

In {\it quantum\/} theories {\it Dirac charge quantization\/} asserts that
these charges are constrained to lie in a lattice in real cohomology.  In
many examples this lattice is the suitably normalized image of integer
cohomology, but recently it was discovered that Ramond-Ramond
charges\footnote{These are often called ``D-brane charges,'' but that is a
misnomer.  After all, in ordinary electromagnetism the notion of charge is
attached to the abelian gauge field, not to the point particles, monopoles,
etc. which are charged with respect to it.  Similarly, Ramond-Ramond fields
have associated charges.  D-branes are Ramond-Ramond charged, just as point
particles are electrically charged.} in Type~II superstring theory lie in the
suitably normalized image of complex $K$-theory instead.  (See~\cite{W3} and
the references contained therein.)  Furthermore, physical arguments suggest
that there is a refined Ramond-Ramond charge in $K$-theory whose image in
real cohomology is the cohomology class of the Ramond-Ramond current.
Inspired by this example, we argue in~\S{2} that the group of charges
associated to {\it any\/} abelian gauge field is a generalized cohomology
group.  The rationale is that the group of charges attached to a manifold~$X$
should depend locally on~$X$, and generalized cohomology groups are more or
less characterized as being topological invariants which satisfy locality (in
the form of the Mayer-Vietoris property).  The choice of generalized
cohomology theory and its embedding into real cohomology affect both the
lattice of charges measured by the gauge field and also the possible torsion
charges.  Both integral cohomology and $K$-theory (in many of its variations)
occur in examples; I do not know an argument to rule out more exotic
cohomology theories.  Particular physical properties---decay processes,
anomalies, etc.---are used to determine which generalized cohomology theory
applies to a particular gauge field.  We do not review such arguments in this
paper.
 
Our concern instead is a more formal question: How do we implement
generalized Dirac charge quantization in a functional integral formulation of
the quantum theory?  The quantization of charge means that the currents have
the local degrees of freedom of a differential form yet carry a global
characteristic class in a generalized cohomology theory.  Furthermore, the
fact that currents and gauge fields couple means that gauge fields are the
same species of geometric object.  We answer this query using {\it
generalized differential cohomology theories\/}.  The marriage of integral
cohomology and differential forms, which we term ordinary differential
cohomology, appears in the mathematics literature in two guises: as
Cheeger-Simons differential characters~\cite{CS} and as smooth Deligne
cohomology~\cite{D}.  For field theory we must go beyond differential
cohomology groups and use cochains and cocycles.  Again this is due to
locality---gauge fields have automorphisms (gauge transformations) and we
cannot cut and paste equivalence classes.  For more subtle reasons electric
and magnetic currents must also be refined to cocycles.  Many aspects of a
cocycle theory for ordinary differential cohomology are developed
in~\cite{HS}, and for generalized cohomology theories it is an ongoing
project of the author, M\. Hopkins, and I\. M\. Singer.  That theory is the
mathematical foundation for the discussion in this paper; we give a
provisional summary in~\S{1}.  The application to abelian gauge theory is one
motivation for the development of generalized differential cohomology theory,
and indeed the presentation here will help shape the theory.  There are other
mathematical motivations for generalized differential cohomology as well.
 
The heart of the paper is~\S{2}, where we write the action for an abelian
gauge field in the language of generalized differential cohomology.  Both
electric and magnetic currents are cocycles for a differential cohomology
class.  The gauge field is a cochain which trivializes the magnetic current;
this is a geometric version of the Maxwell equation~$dF=j_B$.  The electric
current appears in the action; in classical electromagnetism the other
Maxwell equation is the Euler-Lagrange equation.  That term in the action is
anomalous if there is both electric and magnetic current, and the anomaly has
a natural expression~\thetag{2.30} in the language of differential
cohomology.  It is a bilinear form in the electric and magnetic
currents~$j_E$ and~$j_B$.  The various ingredients which enter the discussion
are collected in \theprotag{2.32} {Summary}.  We also describe how twistings
of differential cohomology enter; they are closely related to orientation
issues.
 
Our illustrations in~\S{2} are mostly for 0-form and 1-form gauge fields.
In~\S{3} we turn to theories of more current interest, where higher degree
gauge fields occur.  After a brief comment on Chern-Simons theory, we focus
on superstring theories in 10~dimensions.  There is a new theoretical
ingredient: {\it self-dual\/} gauge fields.  In \theprotag{3.11} {Definition}
we specify the additional data we need to define a self-duality constraint.
The main ingredient is a quadratic form, whose use in defining the partition
function of a self-dual field was elucidated in~\cite{W2}.  Here we also
observe that the same quadratic form is used to divide the usual electric
coupling term by two; see~\thetag{3.26} for the action of a self-dual gauge
field.  Therefore, the quadratic form enters into the formula for the anomaly
as well.  Note that for self-dual fields the electric and magnetic currents
are (essentially) equal.  In this paper we do not explain how the data which
define the self-duality constraint are used in the quantum theory.  These
ideas were applied to D-branes in Type~II superstring theory in~\cite{FH},
where the focus is anomaly cancellation.  We review that argument briefly
here.  For the theory with nonzero Neveu-Schwarz $B$-field twistings of
generalized cohomology play a crucial role.  Indeed, the Ramond-Ramond fields
are cochains in $B$-twisted differential $K$-theory.  In this language a
certain restriction on D-branes (equation~\thetag{3.34}) appears naturally.
We also explain a puzzle~\cite{BDS} about the formula for Ramond-Ramond
charge with nonzero $B$-field.
 
At the end of~\S{3} we treat the Green-Schwarz anomaly cancellation in the
low energy limit of Type~I superstring theory, including global anomalies.
(As far as we know these global anomalies have not previously been
discussed.)  Since the charges in Type~I have been shown to live in
$KO$-theory, the 2-form gauge field is naturally interpreted in differential
$KO$-theory.  As with Type~II the formulation is self-dual.  But here there
are background electric and magnetic currents which are present even in the
absence of D-branes.  Their presence is most naturally explained in our
framework by the observation that the $KO$~quadratic form which defines the
self-duality constraint is not symmetric about the origin.  Rather, the
center is a differential $KO$~ class which determines the background charges.
The theory of this center is discussed in Appendix~B, written jointly with
M. Hopkins.  The gauge field in Type~I is a trivialization of the background
magnetic current, and this leads to a constraint~\thetag{3.46} in $KO$-theory
which generalizes the usual cohomological constraint.\footnote{Let $E$~be the
rank~32 real bundle over spacetime~$X^{10}$.  The usual constraint asserts
that both $X$~and $E$~are spin, and that $\lambda (E)=\lambda (X)$, where
$2\lambda =p_1$.} For spacetimes of the form Minkowski spacetime cross a
compact $r$-dimensional manifold the $KO$~constraint is no new information
if~$r\le 7$.  The computational aspects of the anomaly cancellation in our
treatment are not different than the original, though novel computations are
required to relate our self-dual $KO$-formulation with the standard
formulation in terms of a 2-form field.  We also verify the local and global
anomaly cancellation for D1-~and D5-branes.  The case of Type~I theories
``without vector structure'' also fits naturally into our approach---it
involves a twisted version of~$KO$---but we do not develop the underlying
mathematical ideas.  These global anomaly cancellations are further evidence
that $KO$-theory is the correct generalized cohomology theory for the gauge
field in Type~I.
 
The Atiyah-Singer index theorem, in a geometric form, computes the pfaffian
of a Dirac operator as an integral in differential $KO$-, $KSp$-, or
$K$-theory, depending on the dimension.  In quantum field theories it appears
as the anomaly of the fermionic functional integral.  Sometimes these fermion
anomalies cancel among each other.  In the Green-Schwarz mechanism these
fermion anomalies cancel against a boson anomaly: the anomaly in the electric
coupling of an abelian gauge field in the presence of nonzero magnetic
current.  The gauge field is quantized by some flavor of $K$-theory, and so
the anomaly in the electric coupling is also an integral in a version of
differential $K$-theory.  This idea was first presented in~\cite{FH}.  It
indicates that gauge fields involved in this type of anomaly cancellation
will always be quantized by some variation of $K$-theory.
 
Each factor in an exponentiated (effective) action is a section of a complex
line bundle with metric and connection.  That geometric line bundle is called
the ``anomaly,'' and to say the anomaly cancels between two factors is to say
that the tensor product of the corresponding geometric line bundles is
isomorphic to the trivial bundle.  To define the product of those factors as
a function---so to define the partition function---one needs a choice of
isomorphism.  In this paper we do not address the construction of such
isomorphisms.  It is undoubtedly true that the geometric form of the index
theorem gives a canonical isomorphism between the pfaffian line bundle of a
family of Dirac operators and an integral in some differential $K$-theory.
The definition of the partition function in cases where the Green-Schwarz
mechanism operates depends on this.\footnote{I thank Ed Witten for
emphasizing this point.}

There are two appendices.  The first is a heuristic discussion of Wick
rotation.  We include it since some elementary points, especially in the
context of self-dual gauge fields, cause confusion.  As mentioned above, the
second (with M. Hopkins) contains mathematical arguments needed for the
anomaly cancellation in Type~I.

It is a pleasure to dedicate this paper to Michael Atiyah, Raoul Bott, Fritz
Hirzebruch, and Is Singer.  I hope they enjoy seeing the full-blown
$K$-theory form of the index theorem for families of Dirac operators appear
in physics.  Discussions with many mathematicians and physicists over a long
period of time contributed to the presentation here.  I particularly thank
Jacques Distler and Willy Fischler for clarifying many aspects and Mike
Hopkins for his collaboration on a variety of topological issues in the main
text and in Appendix~B.

 \newpage
 \head
 \S{1} Generalized Differential Cohomology
 \endhead
 \comment
 lasteqno 1@ 18
 \endcomment

In differential geometry we encounter the {\it real\/} cohomology of a
manifold via representative closed differential forms.  In this section we
describe differential geometric objects which represent {\it integral\/}
generalized cohomology classes.  For example, a principal circle bundle with
connection is a differential geometric representative of a degree two
integral cohomology class.  A detailed development of the ideas outlined here
is the subject of ongoing work with M\. Hopkins and I\. M\. Singer.  The
treatment here is only a sketch, offered as background for the discussion of
abelian gauge fields in~\S{2}.
 
Let $\Gamma$ ~be a multiplicative generalized cohomology theory.\footnote{For
simplicity of exposition we assume that the cohomology theory~$\Gamma$ is
multiplicative---all of our examples are---but much of what we say does not
require this hypothesis.} We give examples shortly, but in brief
$\Gamma$~obeys the axioms of ordinary cohomology~$H$ except that the
ring\footnote{Throughout, $A^\bul$ denotes a $\ZZ$-graded abelian group
$A^\bul=\oplus _{q\in \ZZ}A^q$.  Often it has a graded ring structure as
well.  If $A^\bul,B^\bul$~are graded groups, then $A^\bul\otimes B^\bul$~is
double graded.  We denote the associated simply graded group as~$(A\otimes
B)^\bul$.}  $\ring$~ may differ from $H^{\bul}(pt)\cong \ZZ$.  We introduce
the notation
  $$ \pi _{-n}\Gamma = \Gamma ^0(S^{-n}) = \Gamma ^n(pt),\qquad n\in \ZZ, $$
for the cohomology of a point.  (Another typical notation for this graded
ring is~`$\gcb$'.)  The most important property of a generalized cohomology
theory is the Mayer-Vietoris exact sequence, which we view as asserting the
locality of the assignment $X\mapsto \gcb(X)$, where $X$~ranges over a
suitable category of finite dimensional spaces.  Now after tensoring with the
reals, $\Gamma$~is isomorphic to ordinary real cohomology.  More precisely,
there is for each~$X$ a {\it canonical\/} map
  $$ \aligned
      \gcb(X)&\longrightarrow \bigl(H(X;\RR)\otimes \Gamma (pt) \bigr)^\bul
     \\
      \lambda\quad & \longmapsto\quad \quad \lambda _{\RR}
     \endaligned\tag{1.1} $$
It is natural to introduce the notation $\pi _{-n}\Gamma
_{\RR}=\rring^n(pt)=\Gamma ^n(pt)\otimes \RR$.  Then the codomain
of~\thetag{1.1} is the (hyper)cohomology of~$X$ with coefficients in the
graded ring~$\pbring$.  The image of~\thetag{1.1} is a full lattice
$\gbb(X)\subset H\bigl(X;\pring \bigr)^{\bul}$; the kernel is the torsion
subgroup of~$\gcb(X)$.

        \example{\protag{1.2 (integral cohomology)} {Example}}
 There are many cochain models for integral cohomology: singular, \v Cech,
Alexander-Spanier, etc.  Such cochains have integral coefficients, and on the
cochain level the map~\thetag{1.1} is the standard
inclusion~$\ZZ\hookrightarrow \RR$.  A class in the image of~\thetag{1.1} is
represented by a closed differential form~$\omega $ on~$X$ such that
$\int_{Z}\omega $ is an integer for all cycles~$Z$ in~$X$.
        \endexample

        \example{\protag{1.3 ($K$-theory)} {Example}}
 Historically, this is the first example of a generalized cohomology
theory~\cite{BS}, \cite{AH}.  For $X$~compact we can represent an element
of~$K^0(X)$ by a $\zt$-graded vector bundle~$E=E^0\oplus E^1$, thought of as
the formal difference $E^0-E^1$.  The cohomology ring of a point is $\pi
_{-\bul} K\cong \ZZ[[u,u\inv ]]$, where $\deg u=2$.  The element $u\inv \in
K^{-2}(pt)\cong K^0(S^2)$ is called the {\it Bott element\/}; multiplication
by~$u\inv $ is the Bott periodicity map.  The element~$u\inv $ is represented
by the hyperplane (Hopf) complex line bundle over~$\CP^1\cong S^2$.  The
map~\thetag{1.1} is the Chern character
  $$ \ch\: K^\bul(X)\longrightarrow H\bigl(X;\RR[[u,u\inv ]]
     \bigr)^\bul.  $$
        \endexample

        \example{\protag{1.4 ($KO$- and $KSp$-theory)} {Example}}
 These are the variations of $K$-theory for real and quaternionic bundles,
respectively.  Whereas $KO^{\bul}(X)$ is a ring---the tensor product of real
bundles is real---$KSp^{\bul}(X)$~is not.  In fact, $KSp^{\bul}(X)$~is a
module over~$KO^\bul(X)$ and there is also a tensor product
$KSp^{\bul}(X)\otimes KSp^\bul(X)\to KO^\bul(X)$.  So it is natural to
consider the $(\ZZ/2\ZZ\times \ZZ)$-graded theory $KOSp^{\bul}=KO^\bul\times
KSp^\bul$, which does have a multiplicative structure.  Notice that the
ring~$\pi _{-\bul} KOSp$ has torsion in this case.  Over the reals there is an
isomorphism $\pi _{-\bul} KOSp@,@,@,@,@,@,\mstrut _{\RR}\cong \RR[[u^2,u^{-2}]]$.
The element $u^{-2}\in KSp^{-4}(pt)\cong KSp^0(S^4)$ is represented by the
hyperplane (Hopf) quaternionic line bundle over~$\HH\PP^1\cong S^4$.  Odd
powers of~$u^{-2}$ are quaternionic; even powers are real.  Note also that
twice a quaternionic bundle (e.g.~$2u^{-2}$) is real.
        \endexample

{\it Differential $\Gamma $-theory\/}, which we denote~$\Gh$,
combines~$\Gamma$ with closed differential forms~$\clf$.  It is defined on the
category of smooth manifolds.  Loosely speaking, it is the pullback in the
diagram
  $$ \CD 
      \Ghb(\;\cdot\; ) @>>> \clf\bigl(\;\cdot\; ;\pring
     \bigr)^{\bul}\\ 
      @VVV @VVV \\
      \gcb(\;\cdot\; ) @>>> H\bigl(\;\cdot\; ;\pring \bigr)^\bul 
      \endCD \tag{1.5} $$
The northeast corner is the set of closed differential forms with
coefficients in~$\pbring$.  As a first approximation to~$\Gh$, for a
manifold~$X$ define the group~$\agb(X)$ by the pullback diagram 
  $$ \CD 
      \agb(X ) @>>> \clf\bigl(X ;\pring
     \bigr)^{\bul}\\ 
      @VVV @VVV \\
      \gcb(X ) @>>> H\bigl(X ;\pring \bigr)^\bul 
      \endCD   $$
In other words, for each~$q\in \ZZ$ 
  $$ A^q_\Gamma (X) = \Bigl\{\; (\lambda ,\omega )\in \Gamma ^q(X)\times
     \clf(X;\pring )^q : \lambda _{\RR}=[\omega ]_{\text{dR}}\;\Bigr\}.
     \tag{1.6} $$
Here $[\omega ]_{\text{dR}}$ is the de Rham cohomology class of the
form~$\omega $.  But $\Gh$---the pullback in~\thetag{1.5}---is a pullback
{\it as a cohomology theory\/}.\footnote{There is a subtlety which we avoid
in the main text.  Namely, the cohomology theory whose
$q^{\text{th}}$~cohomology is~$\clf^q$ depends on~$q$.  Precisely, on a
manifold~$X$ we use the cochain complex
  $$ \Omega \bigl(X;\pring \bigr)^q\;@>d>>\; \Omega \bigl(X;\pring
     \bigr)^{q+1}\;@>d>>\; \Omega \bigl(X;\pring \bigr)^{q+2} \;@>>>\;\cdots
      $$
to define the theory in the northeast corner of~\thetag{1.5}.  So for
each~$q\in \ZZ$ we have a pullback diagram~\thetag{1.5}.  This leads to a
{\it bi\/}graded cohomology theory in the northwest corner: the
$p^{\text{th}}$~cohomology in the $q^{\text{th}}$~theory is
denoted~$\Ghq q^p$.  The groups we call~$\Gh^q$ are the diagonal
groups~$\Ghq q^q$ in the bigraded theory.}  So a class in~$\Gh^q(X)$ is a
pair~$(\lambda ,\omega )$ with~$\lambda _{\RR}=[\omega ]_{\text{dR}}$ as
in~\thetag{1.6}, together with an ``isomorphism'' of~$\lambda _{\RR}$
and~$[\omega ]_{\text{dR}}$ in~$H\bigl(X;\pring \bigr)^q$.  If we understand
cohomology classes on~$X$ to be homotopy classes of maps from~$X$ into some
universal space~$B$, then an ``isomorphism'' is an explicit choice of
homotopy (up to homotopies of the homotopy).  Even when $\lambda =\omega =0$
there may be nontrivial isomorphisms, and this is the sense in which
$\Gh$~carries topological information beyond~$\Gamma$.  Equivalence
classes of nontrivial isomorphisms appear as the kernel torus in the exact
sequence
  $$ 0 \longrightarrow \frac{H \bigl(X;\pring\bigr)^{q-1}}{\overline{\Gamma}
     ^{q-1}(X)\hphantom{q-}} \longrightarrow \Gh^q(X) \overset c
     \to\longrightarrow A_\Gamma ^q(X) \longrightarrow 0. \tag{1.7} $$
In some situations the kernel torus sometimes captures topological
information not detected by the topological group~$\Gamma ^q(X)$.  If $\lh\in
\Gh^q(X)$ with $c(\lh) = (\lambda ,\omega )$ it is natural to call $\lambda $
the {\it characteristic class\/} of~$\lh$ and $\omega $~the {\it curvature\/}
of~$\lh$.  We can rewrite this exact sequence as
  $$ 0 \longrightarrow \frac{\Omega
     (X;\pring)^{q-1}}{\clf(X;\pring)^{q-1}_\Gamma } \longrightarrow \Gh^q(X)
     \longrightarrow \Gamma ^q(X)\longrightarrow 0, \tag{1.8} $$
where $\clf(X;\pring)^{\bul}_\Gamma $ is the set of closed differential forms
whose cohomology class lies in~$\gbb(X)$.  The second map is the
characteristic class.  The curvature of a differential cohomology class
defined by a global $(q-1)$-form~$B$ is the exact $q$-form~$dB$.  A third way
to present~\thetag{1.7} and~\thetag{1.8} is the exact sequence 
  $$ 0 \longrightarrow \Gamma  ^{q-1}(X;\RR/\ZZ) \longrightarrow \Gh^q(X)
     \longrightarrow \clf(X;\pring)^q_\Gamma \longrightarrow 0.  \tag{1.9}
     $$
The second map is the curvature.  The kernel is the set of ``flat''
differential cohomology classes, an abelian group whose identity component is
the kernel torus in~\thetag{1.7} and whose group of components is the
torsion subgroup of~$\Gamma ^q(X)$.
 
As with topological cohomology theories there are many possible ways to
represent classes in differential cohomology theories.  In computations we
are free to use whichever model is most convenient.  We use the usual
notations $\ChG^{\bul}(X), \ZhG^\bul(X), \BhG^\bul(X)$ for cochains,
cocycles, and coboundaries in a given model for~$\Gh$.  (This is schematic,
as models do not necessarily involve cochain complexes.)  In any model we
construct a category\footnote{We work in the bigraded theory.  Then the
objects in the category form a set, the set of cocycles~$\ZhG(q)^q(X)$.  If
$\ah',\ah\in \ZhG(q)^q(X)$, then a morphism $\bh\:\ah'\to\ah$ is a
cochain~$\bh\in \ChG(q)^{q-1}(X)$ such that $\ah=\ah'+d\bh$
in~$\ZhG(q)^q(X)$, but we take such cochains up to equivalence.  An
equivalence $\chh\:\bh'\to\bh$ is a cochain~$\chh\in \ChG(q)^{q-2}(X)$ with
$\bh=\bh'+d\chh$ in~$\ZhG(q)^{q-1}(X)$.  The group of automorphisms of any
cocycle is $\Gh(q)^{q-1}(X)\cong \Gamma ^{q-1}(X)\otimes \RR/\ZZ$.  The
construction of a category from a cochain complex is standard.  It may be
continued to construct higher categories as well.} whose set of equivalence
classes is~$\Ghb(X)$.  The homotopy theory neatly encodes the categorical
(and multi-categorical) structure in cochain complexes, or better in spaces
of maps.  We need the notion of a ``trivialization'' of a cocycle~$\ah\in
\ZhG^q(X)$.  For our purposes\footnote{There are different notions of
trivialization, and they appear naturally in the bigraded theory.  The most
useful notion makes precise the one mentioned in the text: A trivialization
of a cocycle~$\ah\in \ZhG(q)^q(X) $ is a cochain $\bh\in \ChG(q-1)^{q-1}(X) $
such that $d\bh =\ah $ in~$\ChG(q-1)^q(X)$.  A map $\chh\:\bh'\to\bh$ of
trivializations is then a cochain~$\chh\in \ChG(q-1)^{q-2}(X)$ with
$\bh=\bh'+d\chh$ in~$\ChG(q-1)^{q-1}(X)$.  One can go on to discuss
equivalence classes of such maps and make a category of trivializations
of~$\ah$, analogous to the discussion in the previous footnote.}  we take it
to be a cochain~$\bh\in \ChG^{q-1}(X)$ such that~$d\bh=\ah$.  The meaning
of~`$d$' in this equation depends on the model.  Associated to~$\bh$ is a
differential form~$\eta \in \Omega ^{q-1}(X;\pring)$---the {\it covariant
derivative\/} of~$\bh$---such that $d\eta =\omega $, where $\omega \in
\clf^q(X;\pring)$ is the curvature of~$\ah$.

We next give some explicit models for~$\Gamma = H$ (integral cohomology)
and~$\Gamma=K$ (complex $K$-theory).  The differential cohomology
groups~$\Hh^\bul(X)$ are also known as the groups of {\it Cheeger-Simons
differential characters\/}~\cite{CS} or as the {\it smooth Deligne
cohomology\/} groups~\cite{D}.

        \example{\protag{1.10 (differential cohomology~\cite{HS})} {Example}}
 We represent an element of~$\Hh^q(X)$ by a triple 
  $$ (c,h ,\omega)\in C^q(X;\ZZ) \times C^{q-1}(X;\RR)\times \Omega ^q(X)
      $$
of differentiable singular cochains~\cite{Wa,\S5.31} and differential forms
which satisfy
  $$ \aligned
      \delta c&=0, \\ 
      d\omega &=0, \\ 
      \delta h&=\omega - c_{\RR} .\endaligned  $$
In the last equation we view the differential form~$\omega $ as a singular
cochain by integration over (smooth) chains.  This last equation very
directly expresses the pullback diagram~\thetag{1.5}; $h$~is the isomorphism
of the images of~$c,\omega $ in a set of cochains
representing~$H^\bul(X;\RR)$.  We also have maps of such triples: 
  $$ (s,t)\: (c,h,\omega)\longrightarrow (c',h',\omega '),  $$
where $s\in C^{q-1}(X;\ZZ)$, $t\in C^{q-2}(X;\RR)$, and 
  $$ \aligned
      c' &=c+\delta s, \\ 
      \omega '&=\omega , \\ 
      h' &=h-s_{\RR} - \delta t.\endaligned  $$
In the category representing~$\Hh^q(X)$ we equate maps~$(s,t)$ and~$(s +
\delta e, t - e_\RR - \delta f)$ for $e\in C^{q-2}(X;\ZZ)$, $f\in
C^{q-3}(X;\RR)$.
 
This may be more neatly formulated as a cochain complex whose
$q^{\text{th}}$~cohomology is~$\Hh^q(X)$. 
        \endexample

        \example{\protag{1.11 (differential cohomology)} {Example}}
 Whereas the last model was based on singular cochains, this model is based
on \v Cech theory.  Fix~$q\ge 0$.  Let $\{U_i\}_{i\in I}$ be an open cover
of~$X$ with ordered index set~$I$, and for integers~$r,s$ set
  $$ \check C^{r,s}(X) = \cases 0,&\text{$s<-1$ or $s>q-1$ or $r<0$};
      \\\prod\limits_{i_0<\dots <i_r}C^0(U_{i_0}\cap \dots \cap U_{i_r}\to
     \ZZ),&s=-1;
      \\
      \prod\limits_{i_0<\dots <i_r}{\Omega } ^s(U_{i_0}\cap \dots \cap
     U_{i_r}),&0\le s\le q-1.\endcases $$
Then $\check C^{\bul,\bul}$ is a double complex with the \v Cech
differential~$\delta $ of degree~(1,0) and a differential~$\check{d}$ of
degree~(0,1) defined by
  $$ \check{d} = \cases \text{inclusion} &\text{on $\check
     C^{r,-1}$};\\ 
      d&\text{on $\check C^{r,s}$, $0\le s \le q-1$}.\endcases  $$
The degree~$q-1$ cohomology of the total complex is~$\Hh^q(X)$.  This is the
model which is described, for example, in~\cite{FW,\S6}.  Note the degree
shift in this description.
        \endexample

        \example{\protag{1.12 (differential $K$-theory)} {Example}}
 Here we only discuss models for~$\Kh^0$.  Our first model requires fixing an
infinite dimensional manifold~$B$ whose homotopy type is the classifying
space~$\ZZ\times BU$ of~$K^0$.  There are many possibilities, for example the
space of Fredholm operators on a separable complex Hilbert space.  For
complex $K$-theory we have $\pi _{-\bul} K_{\RR}=K^\bul_{\RR}(pt) \cong
\RR[[u,u\inv ]]$.  Fix a closed differential form $\omega _B\in
\clf(B;\pi K_\RR)^0$ which represents the Chern character of the universal
$K$-theory class on~$B$.  Then
  $$ \omega _B = \omega ^0_B + \omega ^1_Bu\inv  + \omega ^2_Bu^{-2} + \dots
     ,  $$
where $\omega ^i_B\in \Omega ^{2i}(B)$ represents the
$i^{\text{th}}$~universal Chern character class.   A representative of an
element in~$\Kh^0(X)$ is a triple
  $$ (f,\eta ,\omega )\in \Map(X,B)\times \Omega (X;\pi K_\RR)^{-1}\times
     \Omega (X;\pi K_\RR)^0 $$
with 
  $$ \aligned
      d\omega &=0 \\ 
      d\eta &=\omega -f^*\omega _B .\endaligned  $$
Let $\pi \:[0,1]\times X\to X$ be projection.  Then a map of triples is 
  $$ (F,\sigma )\:(f,\eta ,\omega )\longrightarrow (f',\eta ',\omega '),
      $$
where $F\:[0,1]\times X\to B$ and $\sigma \in \Omega (X;\pi K_\RR)^{-2}$
satisfy
  $$ \aligned
      F_0&=f \\ 
      F_1&=f' \\ 
      \omega '&=\omega  \\ 
      \eta '&=\eta  + \pi _*F^*\omega _B + d\sigma .\endaligned 
     $$
There is an equivalence relation on maps~$(F,\sigma )$: the maps~$(F_0,\sigma
_0)$ and~$(F_1,\sigma _1)$ are equivalent if there exists a homotopy
$\tilde{F}\:[0,1]\times [0,1]\times X\to B$ from~$F_0$ to~$F_1$ and a
form~$\phi \in \Omega (X;\pi K_\RR)^{-3}$ such that $\sigma _1 = \sigma _0 +
\Pi _*\tilde{F}^*\omega _B + d\phi $, where $\Pi \:[0,1]\times [0,1]\times X
\to X$ is projection.
        \endexample

        \example{\protag{1.13 (differential $K$-theory)} {Example}}
 Next we give a more geometric picture of elements in~$\Kh^\bul(X)$, though
we do not give a complete ``cochain model'' which computes differential
$K$-theory.  In other words, we do not specify maps between representatives. 
 
Simply stated: A vector bundle $E\to X$ with connection~$\nabla $ represents
an element of~$\Kh^0(X)$.  Certainly $(E,\nabla )$~determines a pair
$(\lambda ,\omega )\in A^0_K(X)$.  Namely, $\lambda \in K^0(X)$ is the
equivalence class of~$E$ and $\omega =\ch(\nabla )$ is the Chern-Weil
representative of the Chern character using the connection~$\nabla $.  To
make contact with our previous model, assume $\omega _B=\ch(\nabla _B)$ is
the Chern character form of a universal vector bundle with
connection~$(E_B,\nabla _B)$ on~$B$.  Choose a classifying map
$\tilde{f}\:E\to E_B$ and let $f\:X\to B$ be the induced map.  Then
$\tilde{f}^*\nabla _B$ is connection on~$E$, and there is a secondary
(Chern-Simons) form $\eta =\eta (\tilde{f}^*\nabla _B,\nabla )$ with $d\eta
=f^*\omega _B-\omega $.  This gives a triple~$(f,\omega ,\eta )$ as in our
previous model.
 
We can also use Quillen's superconnections~\cite{Q} to represent elements
of~$\Kh^0(X)$.  A superconnection on~$E=E^0\oplus E^1$ has a 0-form piece
which is a pair of maps $E^0 \rightleftarrows E^1$.  We can allow~$E^0,E^1$
to be infinite dimensional if we restrict these maps to be Fredholm.  Such
infinite dimensional superconnections play a prominent role in Bismut's
treatment of index theory~\cite{B}.  The expression of that work and of other
geometric developments in index theory in terms of differential $K$-theory is
part of ongoing research.  We learned recently that some versions of
differential $K$-theory and close relatives, together with applications to
index theory, appear in the literature.  See~\cite{L} and the references
therein. 
        \endexample

There are other, more geometric, models for differential cohomology in low
degree.  First, we have 
  $$ \aligned
      \Hh^0(X) &\cong  H^0(X;\ZZ), \\ 
      \Hh^1(X) &\cong  \Map(X,\RR/\ZZ).\endaligned  $$
A circle bundle with connection represents an element of~$\Hh^2(X)$.  There
are various concrete models for elements of~$\Hh^3(X)$, often called ``circle
gerbes with connection''.  See~\cite{Bry}, \cite{H}, \cite{CMW}, \cite{G} for
example.
 
Multiplication and pushforward on~$\Gh$ are induced from the corresponding
operations on~$\Gamma$ and~$\clf$.  Explicit formulas for these operations
depend on the particular cochain model.  Multiplication in~$\Gh$ combines
multiplication in~$\Gamma $ and~$\clf$.  Thus the map~$c$ in~\thetag{1.7} is
a ring homomorphism.  In particular, if $\lh,\lh'$ are differential
cohomology classes with curvatures~$\omega ,\omega '$, and we have a (locally
defined) form~$\alpha $ with~$d\alpha =\omega $, then the curvature of the
product~$\lh\cdot \lh'$ is (locally) the differential of~$\alpha \wedge
\omega $.  Pushforward is defined for suitably oriented maps.  In this paper
we encounter fiber bundles $X\to T$ with compact fibers and inclusions
$i\:W\hookrightarrow X$ of submanifolds.  For a fiber bundle $X\to T$ we need
at least to orient the tangent bundle along the fibers in topological $\Gamma
$-cohomology.  For ordinary cohomology this suffices.  For the various forms
of $K$-theory we also need a Riemannian structure on the family, i.e., a
Riemannian metric on the relative tangent bundle~$T(X/T)$ and a distribution
of horizontal planes on~$X$.  This data determines a Levi-Civita connection
on~$T(X/T)$.  (See~\cite{F2,\S1}.)  We call $X\to T$ a ``Riemannian fiber
bundle.''  Pushforward is integration along the fibers
  $$ \int_{X/T}\:\Ghb(X) \longrightarrow \Gh^{\bul-n}(T), \tag{1.14} $$
where the relative dimension is~$n$.  This map refines to a map on cochain
representatives, and suitable versions of Stokes' theorem hold for this
extension.  For example, if the fibers are closed and a cochain~$\bh $ is a
trivialization of a cocycle~$\ah $, then $\int_{X/T}\bh $~is a trivialization
of~$\int_{X/T}\ah $.  For an inclusion $i\:W\hookrightarrow X$ we must orient
the normal bundle to~$W$ in~$X$ in $\Gamma $-cohomology and also choose a
smooth closed differential form Poincar\'e dual to~$W$.  Then pushforward
  $$ i_*\: \Gh^\bul(W)\longrightarrow \Gh^{\bul+r}(X)  $$
is defined, where $r$~is the codimension of~$W$ in~$X$.  Curvature does not
commute with pushforward.  For example, if $X\to T$ is a $\Gh$-oriented fiber
bundle, then there is a closed differential form~$\AhG(X/T)$ on~$X$ so that
if $\lh\in \Ghb(X)$ has curvature~$\omega $, then the curvature
of~$\int_{X/T}\lh$ is 
  $$ \int_{X/T}\AhG(X/T)\wedge \omega . \tag{1.15} $$
For integral cohomology this form is the constant $\Ahat_H(X/T)\equiv 1$; for
$K$-theory it is $\Ahat(X/T)\wedge e^{\eta /2}$, where $\Ahat$~is the usual
$\Ahat$-genus of the curvature and $-\tp i\eta $ the curvature of a \spinc\
connection on~$X$.  For $KO$- and~$KSp$-theory it is\footnote{Write the
curvature of an element~$\lh\in (\KOh)^0(X)$ as $ \omega =\omega _0 + \omega
_4u^{-2} + \omega _8u^{-4}+\cdots$, where $\omega _i$~are the Chern character
forms of the complexification~$\lh_{\CC}$.  If, for example, $\dim X/T=4$,
then since $\Ahat(X/T)= 1 - p_1(X/T)/24 + \cdots $ we have
  $$ \int_{X/T} \Ahat(X/T)\wedge \omega =\int_{X/T} u^{-2}\bigl(\omega _4
     -\omega _0\,p_1(X/T)/24 \bigr) + \cdots ,  $$
and the curvature of~$\int_{X/T}\lh$ is 
  $$ (2u^{-2})\int_{X/T}\frac 12\bigl(\omega _4
     -\omega _0\,p_1(X/T)/24 \bigr) + \cdots .  $$
Since $2u^{-2}$ is the generator of~$KO^{-4}(pt)\cong \ZZ$, the coefficient
(with the factor~$1/2$) computes the $KO$~index.  }~$\Ahat(X/T)$.
 
Generalized cohomology theories admit twistings, and so too do generalized
differential cohomology theories.  For example, if $F\to X$ is a flat real
vector bundle, then $H^\bul(X;F)$ is a twisted version of real cohomology.
It may be computed by an extension of the de Rham complex to $F$-valued
forms.  Quite generally, for any cohomology theory~$E$ (which could be a
topological theory~$E=\Gamma $ or a differential theory~$E=\Gh$) a real
vector bundle $V\to X$ determines a one-dimensional twisting~$\zeta (V)=\zeta
_E(V)$ of~$E^\bul(X)$.  We denote the $\zeta (V)$-twisted $E$-cohomology
as~$E^{\bul+\zeta (V)}(X)$.  Then there is a Thom homomorphism
  $$ E ^{\bul+\zeta(V) }(X)\longrightarrow E ^{\bul+r}_{\text{cv}}(V),
     \tag{1.16} $$
where $\rank V=r$.  In the codomain we use cohomology with compact vertical
support.  In topological theories \thetag{1.16}~ is an isomorphism, but in
differential theories it only has a left inverse.  For a manifold~$X$ we use
the notation ~$\zeta (X)=\zeta (TX)$ for the twisting derived from its
tangent bundle.  An {\it $E$-orientation\/} of~$V$ is a trivialization of the
twisting~$\zeta (V)$, which then induces an isomorphism~$E ^{\bul+\zeta
(V)}(X)\cong E^\bul(X)$.  For example, in ordinary cohomology $\zeta
(V)=w_1(V)\in H^1(X;\zt)$ is the first Stiefel-Whitney class, the
characteristic class of the real line bundle~$\Det V$.  In topological
$K$-theory $\zeta (V)=\bigl(w_1(V),W_3(V) \bigr)\in H^1(X;\zt)\times
H^3(X;Z)$, so a $K$-theory orientation of~$V$ is an orientation in the usual
sense (trivialization of~$w_1(V)$) together with a $\text{spin}^c$ structure
(trivialization of~$W_3(V)$). In {\it differential\/} $K$-theory the twisting
class is
  $$ \zh (V)=\bigl(w_1(V),\wb_2(V) \bigr)\in H^1(X;\zt)\times \Hh^3(X),
     \tag{1.17} $$
where we use the map $H^2(X;\zt)\to H^2(X;\RR)/H^2(X;\ZZ)\to \Hh^3(X)$
(cf.~\thetag{1.7}) to regard the second Stiefel-Whitney class as a
differential cohomology class (``flat gerbe'') of order two.  A twisting in a
generalized differential cohomology theory induces a twisting of differential
forms.  If $\zh (V)$ is the twisting of a real vector bundle~$V$, the induced
twisting on forms is by the real line bundle $\Det V\to X$.  When $V$~is the
tangent bundle to~$X$, then $\Det V$~is the orientation bundle.  A twisted
$n$-form is simply a density.  (See~\cite{DF,\S2.2} for more about twisted
forms and densities.)
 
Pushforward is defined using the Thom isomorphism, so without choice of
topological orientation makes sense in twisted cohomology.  For example, for
suitable\footnote{For the various forms of $K$-theory we need a Riemannian
fiber bundle; for ordinary cohomology \thetag{1.18}~is true for any fiber
bundle.} fiber bundles $X\to T$ we have
  $$ \int_{X/T}\:\Gh^{\bul-\zh (X/T)}(X) \longrightarrow \Gh^{\bul-n}(T),
     \tag{1.18} $$
where $\zh (X/T) = \zh \bigl(T(X/T) \bigr)$ is the twisting class of the
tangent bundle along the fibers.

 \newpage
 \head
 \S{2} Gauge Fields and Quantization
 \endhead
 \comment
 lasteqno 2@ 36
 \endcomment

In a classical nonrelativistic formulation Maxwell's equations concern a
time-varying electric field $E\in \Omega ^1(\RR^3)$, a time varying magnetic
field~$B\in \Omega ^2(\RR^3)$, a time-varying electric current~$J_E\in \Omega
^2(\RR^3)$, and a time varying electric charge density\footnote{We write
$\rho _E,J_E$ as forms, rather than densities, using the canonical
orientation of~$\RR^3$.  We discuss the role of orientation at the end of
this section.  Using the standard metric and volume form on~$\RR^3$ as well,
we can write~$E,B,J_E$ as vector fields and $\rho _E$~as a function.}~$\rho
_E\in \Omega ^3(\RR^3)$.  The relativistic invariance is manifest if we work
instead on Minkowski spacetime~$M^4=\RR_{@,@,@,@,@,t}\times \RR^3$, where
$t$~is the time coordinate and the speed of light has been set to unity.
Introduce
  $$ \alignedat2
      F &:= B - dt\wedge E\qquad &&\in \Omega ^2(M^4), \\ 
      j_E &:= \rho _E - dt\wedge J_E \qquad &&\in \Omega ^3(M^4).\endaligned
      $$
Then Maxwell's equations assert 
  $$ \aligned
      dF &= 0, \\ 
      d*F &= j_E.\endaligned \tag{2.1} $$
With an eye towards generalizations we introduce a magnetic current~$j_B\in
\Omega ^3(M^4)$ and allow~$dF$ to be nonzero:
  $$ \aligned
      dF &= j_B, \\ 
      d*F &= j_E.\endaligned \tag{2.2} $$
This version of Maxwell's equations is our starting point.
 
The form~$F$ is called the {\it field strength\/}, $j_E$~the {\it electric
current\/}, and $j_B$~the {\it magnetic current\/}.  We assume that on any
spacelike slice $j_E,j_B$~have compact support (or more generally satisfy
some integrability condition).  The integral of~$j_E$ ($j_B$) over a
spacelike slice~$N\cong \RR^3$ is the total {\it electric\/} ({\it
magnetic\/}) {\it charge\/}.  Maxwell's equations~\thetag{2.2} imply that the
currents~$j_E,j_B$ are closed, and as a consequence the total charge is
constant in time.  Letting~$j$ denote either of the closed forms~$j_E$
or~$j_B$ we have a de Rham cohomology interpretation of the charges~$\Qb _E$
and~$\Qb _B$:
  $$ \Qb  = \left[ j\res N \right]_{dR}\in H^3_c(N;\RR).  \tag{2.3} $$
The subscript~`$c$' indicates that the cohomology is taken with compact
support.  Notice that the field strength~$F$ need not have compact support,
so equation~\thetag{2.2} does not imply that the charge vanishes.  However,
it does imply
  $$ \Qb  \in \ker\bigl(H_c^3(N;\RR)\longrightarrow H^3(N;\RR) \bigr). \tag{2.4}
     $$
When $N\cong \RR^3$ this refinement is vacuous, but for more general
manifolds~$N$ (of higher dimension, for example) it may be nontrivial.  For
example, if $N$~is compact \thetag{2.4}~implies that~$\Qb =0$.  For a similar
discussion of charge, see~\cite{MW,\S2}.
 
So far we have presented the equations of {\it classical\/} electromagnetism.
A new feature enters in the {\it quantum\/} theory: charge is quantized.
{\it Dirac charge quantization\/} asserts that in appropriate units the total
charge is an integer.  Equivalently, the cohomology class~$\Qb $
in~\thetag{2.3}, \thetag{2.4} lies in the image of the map $H_c^3(N;\ZZ)\to
H^3_c(N;\RR)$.  This is the correct quantization condition for Maxwell
theory.  For general abelian gauge fields integral cohomology may be replaced
by a generalized cohomology theory, as we now explain.
 
We work in arbitrary dimensions and allow space~$N$ to be any oriented
Riemannian manifold of dimension~$n-1$. Let~$F$ be an (abelian) field
strength and $j_E,j_B$~currents.  These are real differential forms
on~$\RR\times N$.  We allow~$F$ to have arbitrary degree; it may or may not
be homogeneous.  (Shortly we will consider~$F$ as a form with coefficients,
as in~\S{1}.)  The degrees of the currents are then determined
from~\thetag{2.2}.  The currents are assumed to have compact support on
spacelike slices, but we do not make the support condition explicit in the
notation.  Quantization of the charge associated to~$F$ means that the
integral of a current~$j$ over a closed cycle in~$N$ is not an arbitrary real
number, but rather takes discrete values.  Therefore, the charge $\Qb
=[j]$~is restricted to lie in a lattice $\overline{\Gamma }^\bul(N)\subset
H^\bul(N;\RR)$.  It is natural from a mathematical point of view---there are
physical arguments which motivate this---to postulate an abelian
group~$\gcb(N)$ and a map $\gcb(N)\to H^\bul(N;\RR)$ with image~$\gbb(N)$
such that the charges~$\Qb \in \gbb(N)$ are refined to charges~$Q\in
\gcb(N)$.  Furthermore, the locality of quantum field theory implies that the
group of possible charges~$\gcb(N)$ should depend locally on~$N$.  As stated
at the beginning of~\S{1}, locality is a characteristic feature of
generalized cohomology theories whose expression is the Mayer-Vietoris exact
sequence.  We are led, then, to postulate that the group of charges~$\gcb(N)$
assigned to a space~$N$ is a generalized cohomology group.  We will not
discuss the physical motivation behind the choice of~$\Gamma $ and the choice
of map to real cohomology.  There are detailed discussions of particular
cases in the physics literature (most recently concerning Ramond-Ramond
fields in Type~II superstring theory and its close cousins).  Notice that
different choices of~$\Gamma $ and the map to real cohomology lead to
different lattices~$\gbb(N)$, so to different quantization conditions on
charges measured around cycles.  Also, different choices of~$\Gamma $ lead to
different torsion phenomena for charges.

Now we Wick rotate to Euclidean field theory and formulate the theory on
oriented\footnote{At the end of this section we relax the orientation
assumption.} Riemannian manifolds~$X$ of dimension~$n$.  (Appendix~A reviews
Wick rotation in general terms, so provides the setting for our discussion
here.)  We will not specify explicit support conditions on currents, though
the reader should keep in mind the compact spatial support condition above on
manifolds of the form $X=\RR \times N$.  Correlation functions in the
Euclidean theory are defined (formally) by a functional integral over
Euclidean fields using a Euclidean action.  Our task is to describe the
Euclidean fields and Euclidean action precisely.  We implement our conclusion
in the previous paragraph in the Euclidean setting by choosing: (i)\ a
generalized cohomology theory~$\Gamma $, and (ii)\ a map
  $$ \gcb(X)\to H(X;\pring)^\bul \tag{2.5} $$
to real cohomology.  This is precisely the data we need to define
differential $\Gamma $-theory~$\Gh$.  Now given a generalized cohomology
theory~$\Gamma $ there is a canonical map~\thetag{1.1}, and any other
map~\thetag{2.5} is obtained by multiplying by an invertible element
in~$H(X;\pring)^0$.  In gauge theory there is an invertible closed
differential form
  $$ \omega _X\in \clf(X;\pring)^0 \tag{2.6} $$
which represents this class; it depends locally on~$X$ in a suitable sense.
Then as we will see shortly, it is then natural to lift the currents~$j$ to
differential cohomology classes~$\jh\in \Ghb(X)$ whose image in~$\agb(X)$
is\footnote{We continue to use ``charges''~$Q,\Qb$ in the Euclidean setting.
The physical interpretation of these quantities as charges is in the
Hamiltonian situation $X=\RR\times N$ after rotating back to real time.}
  $$ c(\jh) = (Q,\frac{j}{\omega _X}). \tag{2.7} $$
(Recall the definition of~$\agb$ from~\thetag{1.6}.)  Note that $\Qb$~is
$[\omega _X]_{dR}$~times the image of~$Q$ under~\thetag{2.5}.  A particularly
good choice for the normalizing factor~\thetag{2.6} is
  $$ \omega _X = \tp\sqrt{\AhG(X)}, \tag{2.8} $$
where $\AhG(X)$ is defined in~\thetag{1.15}.  (The~$\tp$ is convention; the
$\sqrt{\AhG(X)}$ is to make bilinear pairings in~$\Gamma $ compatible with
integration of curvatures.)  To make sense of the first Maxwell
equation~$dF=j_B$ we must refine the field strength~$F$ to differential
$\Gamma $-theory as well.  We will see that in fact we must lift~$F$ and~$j$
to {\it cocycles\/} representing generalized differential cohomology classes.
(If $j_B\not= 0$ then $F$~is lifted to a cochain rather than a cocycle.)  In
this setting the differential forms $F$~and $j$~have coefficients
in~$\pbring$.  We proceed with the construction of the Euclidean theory after
describing two motivating examples.

        \example{\protag{2.9 (typical $p$-form gauge field)} {Example}}
 Suppose $F$~is a homogeneous form of degree~$p+1$, so that $\deg j_B=p+2$
and $\deg j_E = n-p$.  Typically the quantization law asserts that charges
lie in integral cohomology~$\Gamma =H$ with $\omega _X=\tp$.  In other words,
the pair~$(Q_E,Q_B)$ lives in~$H^{p+2}(X;\ZZ)\oplus H^{n-p}(X;\ZZ)$.
        \endexample

        \example{\protag{2.10 (Ramond-Ramond fields)} {Example}}
 These occur in the low energy field theory description of the Type~II
superstring.  Here $X^{10}$~is a spin Riemannian manifold.  If the $B$-field
vanishes, then the Ramond-Ramond charges naturally live in $K$-theory.  (For
a physical discussion of the choice of $K$-theory, see~\cite{W3} and the
references therein.)  The charge is a homogeneous class in~$K^\bul(X)$, and
by Bott periodicity only the parity of the degree matters.  The parity is odd
in Type~IIA and even in Type~IIB.  For definiteness we suppose the charge
lives in~$K^1(X)$ in Type~IIA and $K^0(X)$~in Type~IIB.  (For Type~IIB---and
probably for any theory---the charge is in~$\epsilon \inv (0)$ for $\epsilon
\:K^0(X)\to H^0(X)$ the augmentation.)  Thus the Ramond-Ramond field
strengths and currents are refined to differential $K$-theory classes of
degree~0 and~$-1$, respectively.  In this case \thetag{2.8}~is~$\omega
_X=\tp\sqrt{\Ahat(X)}$.  Also, there is a self-duality condition which enters
in the construction of the functional integral.  We discuss self-dual fields
in~\S{3}.
 
The $B$-field, which heretofore was assumed zero, is locally a 2-form.  Its
field strength, usually denoted~$H$, is a closed 3-form on~$X$ which obeys
integrality constraints corresponding to integral cohomology.  In other
words, we postulate a class~$\zeta \in H^3(X)$ with $\zeta _\RR = [H]_{dR}$,
and suppose that globally the $B$-field is a cocycle~$\zh$ for a class in the
differential cohomology group~$\Hh^3(X)$.  Then the Ramond-Ramond charges
live in twisted $K$-theory~$K^{\bul+\zeta}(X)$, and currents are lifted to
twisted differential $K$-theory~$\Kh^{\bul+\zh}(X)$.
        \endexample

There is a lagrangian formulation of the classical Maxwell
equations~\thetag{2.1} (with no magnetic current) in the classical Lorentzian
theory.  The field variable is a gauge field.  The first Maxwell equation is
the Bianchi identity and holds off-shell.  The second Maxwell equation is the
variational equation of a classical action for the gauge field.  As stated
above, our task is to incorporate Dirac charge quantization into the Wick
rotated Euclidean theory.  (We summarize our answer in~\theprotag{2.32}
{Summary}.)  As we have seen charge quantization means choosing a generalized
cohomology theory~$\Gamma $ and an embedding~\thetag{2.5}, refined to a
differential form~\thetag{2.6}.  We begin with the case where the
currents~$j_E$ and~$j_B$ both vanish. Fix a degree\footnote{We use a
multi-index notation
  $$ \dd = (d_1,\dots ,d_k).  $$
Then if $A^\bul$~is a graded group, an element~$a\in A^{\dd}$ is a $k$-tuple
$(a_1,\dots ,a_k)$ with $a_i\in A^{d_i}$.  Arithmetic is done componentwise.
For example, $\dd + 1 = (d_1+1,\dots ,d_k+1)$.}~$\dd$.  Then we: (i)\ refine
the characteristic class~$[F/\omega _X]_{dR}$ of the normalized field
strength~$F\in \clf(X;\pring)^\dd$ to a class~$\lambda \in \Gamma ^\dd(X)$,
and (ii)\ refine the field strength itself to a cohomology class~$\Fh\in
\Gh^\dd(X)$ such that
  $$ c(\Fh) = \Bigl(\lambda ,\frac{F}{\omega _X} \Bigr).  $$
Here we use a multi-normalization $\omega _X = \bigl((\omega _X)_{1},\dots
,(\omega _X)_{k} \bigr)$ corresponding to the components of $F=(F_1,\dots
,F_k)$.  The differential cohomology group~$\Gh^\dd(X)$ is the space of
abelian gauge fields (or gauge potentials) {\it up to gauge
transformations\/}.  It is the space over which one integrates in the
Euclidean functional integral.  Cocycles in~$\ZhG^\dd(X)$ representing a
class~$\Fh\in \Gh^\dd(X)$ are particular {\it gauge fields\/}.  If $\Ah \in
\ZhG^\dd(X)$ is such a cocycle, we denote its cohomology class
in~$\Gh^{\dd}(X)$ as~$\Fh_A$.  Cocycles are the proper variables for field
theory---they are local.  We are led, then, to a Euclidean theory in which
the space of fields is the category~$\ZhG^\dd(X)$ of cocycles (of particular
degrees).  The Wick rotated version of the classical Lorentzian action makes
sense for our refined fields: If $\Ah \in \ZhG^\dd(X)$ is a field with
curvature~$F_A/\omega _X$, then the Euclidean action is
  $$ S(\Ah ) = \frac 1{2e^2}\int_{X}F_A\wedge *F_A. \tag{2.11} $$
Here $e=(e_1,\dots ,e_k)$ is a set of coupling constants, and the notation
implies a sum over components: 
  $$ S(\Ah) = \sum\limits_{i=1}^k \frac 1{2e_i^2}\int_{X}(F_A)_i\wedge
     *(F_A)_i.  $$
Since the curvature depends only on the cohomology class~$\Fh_A$ of~$\Ah $
in~$\Gh^\dd(X)$, the action is gauge-invariant.  Of course, in
writing~\thetag{2.11} we are implicitly assuming either that $X$~is compact
or some support condition on the fields.

        \example{\protag{2.12 (1-form gauge field)} {Example}}
 Let $X^n$~be an oriented $n$-manifold and suppose~$F\in \clf^2(X)$ obeys
quantization using integral cohomology (with $\omega _X=\tp$).  As a model
for~$\check{Z}_H^2(X)$ we take the category of connections on principal
circle bundles over~$X$.  A gauge field~$\Ah $ is such a connection
and~$\Fh_A$ its equivalence class under isomorphisms of circle bundles with
connection.  The field strength~$F_A$ is $\sqo$~times the curvature of~$\Ah
$; the characteristic class is the first Chern class.  If the characteristic
class vanishes, then the gauge field may be represented by a global 1-form,
uniquely up to differentials of circle-valued functions (see~\thetag{1.8}).
        \endexample

        \example{\protag{2.13 (periodic scalar field)} {Example}}
 Again $X^n$~is an oriented $n$-manifold.  Suppose $F\in \clf^1(X)$, with
quantization specified by integer cohomology.  Now $\check{Z}_H^1(X)$ may
simply be taken to be the {\it set\/}~$\Map(X,\RR/\ZZ)$ of periodic real
scalar fields on~$X$.  Taking~$\omega _X=\tp$ a gauge field is a map $\phi
\:X\to \RR/2\pi \ZZ$ and its field strength is~$d\phi $.  Then \thetag{2.11}~
is the usual action~$\frac 1{2e^2}\int_{X}d\phi \wedge *d\phi $.  It is
convenient in this case to write formulas in terms of the exponentiated
circle-valued scalar field $e^{i\phi }\:X\to\TT$.
        \endexample
 
Next, we allow the currents to be nonzero.  First, consider~$j_E\not= 0$.  In
the classical Lorentzian theory there is an additional term in the action
whose variation gives the right-hand side of the second Maxwell
equation\footnote{The coupling constant is $e^2=2\pi $ in~\thetag{2.1}.}
in~\thetag{2.1}.  To write its Wick rotation in our framework we need to
postulate maps\footnote{For complex $K$-theory there are natural
``determinant'' maps $K^1\to H^1$ and $K^2\to H^2$.  The map $\pi _{-\bul}
K_\RR\cong \RR[[u,u\inv ]] \to \pi _{-\bul} H_\RR\cong \RR$ sends~$u$ to zero.
We will also make use of the natural ``pfaffian'' map $KSp^2\to
H^2$. (See~\cite{F2,\S3}; note $KSp^2\cong KO^{-2}$ by Bott periodicity.)
The map in~\thetag{2.14} in degree~1 is used in~\thetag{2.17}; that in
degree~2 is used later in~\thetag{2.30}.  For general~$\Gamma$ maps to low
degree cohomology with coefficients exist canonically, using the Postnikov
tower, and with some choice one produces~\thetag{2.14}.}
  $$ \aligned
      \Gamma ^1 &\longrightarrow H^1, \\
      \Gamma ^2 &\longrightarrow H^2, \\
      \pbring &\longrightarrow \pi _{-\bul} H_\RR\cong \RR.\endaligned
     \tag{2.14} $$
Using the pullback square~\thetag{1.5} there are induced maps
  $$ \aligned
      \Gh^1 &\longrightarrow \Hh^1\\
      \Gh^2 &\longrightarrow \Hh^2.\endaligned \tag{2.15} $$
We must also assume that $X$~is $\Gh $-oriented so that integration over~$X$
in $\Gh$-theory is defined.  Recall that the closed differential form~$j_E$
is refined to a differential cohomology class~$\jh_E\in \Gh^{n-\dd+1}(X)$.
(See~\thetag{2.7}.)  The additional term in the Euclidean action is the
purely imaginary expression
  $$ \tp i\int_{X} \jh_E \cdot \Fh_A . \tag{2.16} $$
The product takes place in~$\Ghb(X)$, and the degrees are such that the
integrand is an element of\footnote{If $\dd=(d_1,\dots ,d_k)$ is a
multi-degree, then \thetag{2.16}~is the sum of $k$~terms, one for each
component of~$\jh_E$ and~$\Fh_A$.  For the signs to work out properly, we
should regard~$\jh_E$ as a form twisted by the orientation bundle, so of
degree~$-(d-1)$.  (A density---twisted $n$-form---has degree~0.)  This is
formalized in~\thetag{2.33} below.}~$\Gh^{n+1}(X)$.  Hence the integral lands
in~$\Gh^1(pt)$, and using~\thetag{2.15} we map to $\Hh^1(pt)\cong \RR/\ZZ$.
Therefore, the exponentiated action
  $$ e^{-S}(\Ah ) = \exp\bigl(-\frac 1{2e^2}\int_{X}F_A\wedge *F_A
     \bigr)\,\exp\bigl(-\tp i\int_{X}\jh_E\cdot \Fh_A \bigr) \tag{2.17} $$
is a well-defined complex number.
 
Several comments are in order.  First, it is more illuminating to work with a
family of gauge fields~$\Ah $ parametrized by a manifold~$T$.  Then the
exponentiated action is a function $e^{-S}\:T\to\CC$ and we can use Stokes'
theorem to differentiate the second term of the action.  Also, the fact that
\thetag{2.17}~depends on~$\Ah $ only through~$F_A,\Fh_A$ means that the
exponentiated action~$e^{-S}$ is gauge-invariant.  Finally, for gauge fields
quantized by integer cohomology, as in \theprotag{2.12} {Example} and
\theprotag{2.13} {Example}, the electric coupling term~\thetag{2.16} is
usually written as ``$\frac{i}{\tp}\int_{X}j_E\wedge A$''.  Indeed,
\thetag{2.16}~reduces to this if $F_A=dA$ for some form~$A$; otherwise this
is only valid locally.  The correct global expression is~\thetag{2.16}.

        \example{\protag{2.18 (1-form gauge field)} {Example}}
 Continuing \theprotag{2.12} {Example}, a typical electric current~$j_E$ is
induced from point charges.  For manifolds of the form $\RR\times N$ the
point charges are described by a finite set~$P$ of points in~$N$ with
integers attached.  If the particles are static, then $\RR\times P$~is the
set of their ``worldlines''; if they move the worldline is the graph of a
function $\RR\to N$.  More generally, let $W\subset X$ be a 1-dimensional
oriented submanifold and
  $$ q_E\:W\longrightarrow \ZZ \tag{2.19} $$
a (locally constant) function.\footnote{For manifolds of the form~$\RR\times
N$ we require $W\cap\bigl(\{\tau \}\times N \bigr)$ to be compact for
all~$\tau \in \RR$.  For general~$X$ we do not specify support conditions,
and in any case omit them from the notation as usual.}  Let
$i\:W\hookrightarrow X$ denote the inclusion.  Then the electric charge is
the pushforward of~$q_E$ in cohomology:
  $$ Q_E = i_*q_E\;\in H^{n-1}(X;\ZZ). \tag{2.20} $$ %KEEPTAG 
We can regard~$q_E$ as a class in~$\Hh^0(X)$ (or better as a cocycle
in~$\Zh_H^0(W)$).  Then the refined electric current is the pushforward
of~$q_E$ in differential cohomology:
  $$ \jh_E = i_*q_E\;\in \Hh^{n-1}(X). \tag{2.21} $$
Recall from~\S{1} that this depends on choosing a smooth closed differential
form Poincar\'e dual to~$W$.  Without making any choices one could define a
distributional electric current---a current in the sense of de
Rham---supported on~$W$.  But we prefer to remain in the smooth
category.\footnote{Smoothness plays a greater role when we come to magnetic
currents.  For example, it was a key idea in~\cite{FHMM}. The point is to
avoid illegal products of distributions.} The second factor in the
exponentiated action~\thetag{2.17} may be rewritten as
  $$ \exp\bigl(-\tp i\int_{W}\Fh_A\bigr)^{q_E}. \tag{2.22} $$
This is the product over components of~$W$ of the $q_E^{\text{th}}$~power of
the holonomy of the connection~$A$. 
        \endexample

        \example{\protag{2.23 (periodic scalar field)} {Example}}
 There is an analogous story for the periodic scalar $e^{i\phi} \:X\to\TT$,
continuing \theprotag{2.13} {Example}.  In this case $W$~is 0-dimensional---a
``$(-1)$-brane''---and \thetag{2.19}--\thetag{2.21} hold with $n-1$~replaced
by~$n$.  Expression~\thetag{2.22} becomes a product over the points of~$W$: 
  $$ \prod\limits_{w\in W}e^{-iq_E(w)\phi (w)} \tag{2.24} $$
This factor is often viewed as a local operator inserted into the functional
integral, but in our context it is the electric coupling in the exponentiated
action. 
        \endexample

Next, we consider nonzero magnetic current~$j_B\not= 0$.  As before, we
suppose $j_B$~is refined to a differential cohomology class~$\jh_B\in
\Gh^{\dd+1}(X)$, and assume that we have a fixed cocycle representative, also
denoted~$\jh_B$.  Recall the first Maxwell equation in~\thetag{2.2}.  If the
magnetic current is nonzero, then the field strength is not closed and it
makes no sense to refine it to a differential cohomology class.  Rather, we
postulate that the gauge field~$\Ah $ is a trivialization of the refined
magnetic current~$\jh_B$, in the sense explained before \theprotag{1.10}
{Example}.  This means that~$\Ah \in \ChG^\dd(X)$ with $d\Ah = \jh_B$.  The
equivalence class of~$\Ah $ (under the equivalence of trivializations
of~$\jh_B$ discussed in a footnote preceding \theprotag{1.10} {Example}) is
denoted~$\Fh_A$ as before.  The covariant derivative~$F_A/\omega _X$ of~$\Ah
$ is a global differential form, and~$dF_A=j_B$.  Notice that the existence
of a global trivialization of~$\jh_B$ implies that the magnetic charge~$Q_B$
vanishes in the global cohomology group (no support condition), which is
consistent with~\thetag{2.4}.  Also, if~$\jh_B=0$ we recover the previous
definitions of~$\Ah ,\Fh_A,F_A$.

        \example{\protag{2.25 (periodic scalar field)} {Example}}
 We continue \theprotag{2.23} {Example}.  In this case $\deg j_B=2$ and in
our model for~$\Hh^2$ the refined magnetic current~$\jh_B\in \Hh^2(X)$ is a
principal circle bundle with connection on~$X$ whose curvature is
$\sqo$ times~$j_B$.  The exponentiated gauge field~$e^{i\phi }$ is now
a global section of this bundle, and therefore the bundle is topologically
trivial.  The field strength~$F_A=d\phi $ is the ``covariant derivative'' of
this section, that is, the pullback to~$X$ of the connection form on the
total space of the circle bundle.  Notice that in case~$\jh_B=0$---the
cocycle~$\jh_B$ is the trivial circle bundle with product connection---we
recover our previous description of the gauge field as a map $X\to \TT$.

Specialize now to $n=\dim X = 2$.  Then we can consider a magnetically
charged $(-1)$-brane.  Thus suppose $i\:W\hookrightarrow X$ is the inclusion
of an oriented 0-manifold and 
  $$ q_B\:W\longrightarrow \ZZ, \tag{2.26} $$
which we regard as a class in~$\Hh^0(W)$.  Then we set
  $$ \jh_B = i_*q_B\;\in \Hh^2(X). \tag{2.27} $$
Recall that the pushforward depends on a choice of Poincar\'e dual form,
which in this case is a closed bump 2-form localized near points~$w\in W$ and
whose integral near~$w$ is~$q_B(w)$.  A cocycle representative of the refined
magnetic current is a circle bundle with connection whose curvature is this
Poincar\'e dual form.  This construction should be compared to the
construction in complex geometry of a holomorphic line bundle from a
divisor. 
        \endexample

        \example{\protag{2.28 (1-form gauge field)} {Example}}
 In this case $\jh_B\in \check{Z}_H^3(X)$ is intuitively a ``gerbe with
connection'' and $\Ah $~is a trivialization---a ``translated version'' of a
circle bundle with connection.  The theory outlined in~\S{1} is a natural
home for these notions and their generalizations to higher degrees.
        \endexample

If the electric current~$j_E$ vanishes, then the action~\thetag{2.11} is
well-defined and gauge-invariant.  In case both $j_E$~and $j_B$~are nonzero
we must re-examine the second factor in the exponentiated
action~\thetag{2.17}, whose form does not change, but whose geometric nature
does.  As before it is gauge-invariant, since it only depends on~$\Ah $
through~$\Fh_A$.  In a family of gauge fields parametrized by~$T$, we compute
the action as a function on~$T$.  Suppose now that the refined electric
current~$\jh_E$ has been lifted to a cocycle.  Since $\Fh_A$~is a
trivialization of~$\jh_B$, by Stokes' theorem (see the text
following~\thetag{1.14})
  $$ \text{$\exp\bigl(-\tp i\int_{X/T}\jh_E \cdot \Fh_A\bigr)$ \quad is a
     trivialization of\quad $\exp\bigl(-\tp i\int_{X/T}\jh_E\cdot \jh_B
     \bigr)$.}  \tag{2.29} $$
The degrees work out so that 
  $$ \exp\bigl(-\tp i\int_{X/T}\jh_E\cdot \jh_B \bigr)  \tag{2.30} $$
lives in\footnote{Note we use the degree~2 map in~\thetag{2.14} to
define~\thetag{2.30} as an element in~$\Zh_H^2(T)$, though we do not make it
explicit in the notation.}~$\Zh^2_H(T)$, i.e., is a circle bundle with
connection over~$T$.  Equivalently, the first expression in~\thetag{2.29} is
a section of a hermitian line bundle with connection, as then is the entire
exponentiated action~\thetag{2.17}.  Actions which are sections of hermitian
line bundles with connection are potentially anomalous; the anomaly is the
obstruction to trivializing the line bundle with connection.  Here
\thetag{2.30}~is the formula for the anomaly, where we interpret the integral
as a differential cohomology class in~$\Hh^2_H(T)$.

        \example{\protag{2.31 (periodic scalar field)} {Example}}
 We continue \theprotag{2.25} {Example} in any dimension~$n$.  Recall that
$i\:W\hookrightarrow X$ is the inclusion of a 0-manifold---which we assume to
be compact, i.e., a finite set of points---and $q_E\:W\to\ZZ$ encodes the
electric charges of the points of~$W$.  The refined magnetic current
$\jh_B=i_*q_B$ is a circle bundle with connection over~$X$ and the gauge
field~$e^{i\phi }$ is a section of~$\jh_B$.  Let $L\to X$ denote the
associated hermitian line bundle with connection.  The electric coupling in
the exponentiated action is~\thetag{2.24}, and the anomaly formula reduces to
the obvious assertion that it is an element of the hermitian line
  $$ \bigotimes\limits_{w\in W}(L_w)^{\otimes (-q_E(w) )}. 
     $$
As we vary over a family~$T$ of connections (and also
embeddings~$W\hookrightarrow X$) these lines assemble into a smooth hermitian
line bundle with connection over~$T$. 
        \endexample

\flushpar
 The anomaly~\thetag{2.30} is nonzero in this example since $W$~is both
electrically and magnetically charged.  The example generalizes to higher
dimensional submanifolds which are both electrically and magnetically
charged.  In these higher dimensional cases the Euler class~$\chi _{\Gh}(\nu
)$ in $\Gh$-theory of the normal bundle~$\nu $ to~$W$ in~$X$ enters. 
 
For convenience we collate the various parts of the discussion.

        \definition{\protag{2.32} {Summary}}
 The data needed to define an abelian gauge field is: \medskip
 {\narrower\narrower
 \item{(i)} a generalized cohomology theory~$\Gamma $;
 \item{(ii)} maps~\thetag{2.14} to ordinary cohomology; 
 \item{(iii)} a multidegree~$\dd=(d_1,\dots ,d_k)$;
 \item{(iv)} normalizing differential forms $\omega _X =
\bigl((\omega _X)_{1},\dots ,(\omega _X)_{k} \bigr)$ which depend
functorially and locally on~$X$; and
 \item{(v)} coupling constants~$e=(e_1,\dots ,e_k)$.
\par}
 \medskip\flushpar
 Then the gauge field, magnetic current, and electric current live in:
  $$ \aligned
      \Ah &\in \Ch^{\dd}_{\Gamma }(X), \\ 
      \jh _B &\in \Zh^{\dd+1}_{\Gamma }(X), \\ 
      \jh _E &\in \Zh^{n-\dd+1}_{\Gamma }(X).\endaligned
       $$
The gauge field~$\Ah$ is a nonflat trivialization of the magnetic
current~$\jh_B$.  The exponentiated action is~\thetag{2.17}, and the electric
coupling has an anomaly given by~\thetag{2.30}.
        \enddefinition

Finally, we relax the orientability assumption on~$X$.  For this we use the
discussion of twistings and orientation at the end of~\S{1}.  Thus let $X$ be
a Riemannian manifold which is not oriented and possibly not orientable.  The
only change\footnote{There is another possible scenario in which the gauge
field is twisted, hence the magnetic current is twisted, and the electric
current is untwisted.  This occurs in M-theory, for example.} from
\theprotag{2.32} {Summary} is that the refined electric current~$\jh_E$ lives
in~$-\zh (X)$-twisted $\Gh$-theory:
  $$ \jh_E\in \Zh^{n-\dd+1-\zh(X)}_{\Gamma }(X). \tag{2.33} $$
The twisting refers to differential $\Gh$-theory.  Then the integral
in~\thetag{2.16} is well-defined (cf.~\thetag{1.18}).  The refined magnetic
current~$\jh_B$ still lives in the untwisted theory.  Suppose these currents
are induced from submanifolds $i\:W\hookrightarrow X$ and
cocycles~$\qh_E,\qh_B$ on~$W$, as in \thetag{2.19} and~\thetag{2.27}, but now
we allow twisted cocycles.  In other words, postulate twistings $\th _E,\th
_B$ on~$W$ such that
  $$ \aligned
      \qh_E&\in \Zh^{\bul+\th_E}_{\Gamma }(W), \\ 
      \qh_B&\in \Zh^{\bul+\th_B}_{\Gamma }(W).\endaligned  $$
Then if $\jh_E=i_*\qh_E$ and $\jh_B=i_*\qh_B$ the twistings must satisfy  
  $$ \aligned
      \th_E &= \zh(\nu ) - i^*\zh(X),\\ 
      \th_B &= \zh(\nu ), \endaligned\tag{2.34} $$
where $\nu \to W$ is the normal bundle to~$W$ in~$X$.   
 
In some theories the gauge fields and currents live in twisted versions of
differential cohomology.  Precisely, we have for some twisting~$\zh$: 
  $$ \aligned
      \Ah &\in \Ch^{\dd+\zh}_{\Gamma }(X), \\ 
      \jh _B &\in \Zh^{\dd+1+\zh}_{\Gamma }(X), \\ 
      \jh _E &\in \Zh^{n-\dd+1-\zh-\zh(X)}_{\Gamma }(X).\endaligned
     \tag{2.35} $$
Note the sign change in the twisting for the electric current.  This makes
the electric coupling~\thetag{2.16} well-defined.  Equations~\thetag{2.34}
are now 
  $$ \aligned
      \th_E &= \zh(\nu ) + i^*\zh - i^*\zh(X),\\ 
      \th_B &= \zh(\nu ) + i^*\zh. \endaligned  \tag{2.36} $$

 \newpage
 \head
 \S{3} Applications
 \endhead
 \comment
 lasteqno 3@ 55
 \endcomment

 \subhead Chern-Simons Class
 \endsubhead

Recall that the origin of differential cohomology lies in the work of
Cheeger-Simons~\cite{CS}.  Their primary motivation is the application to
secondary characteristic classes.  We focus on 4-dimensional characteristic
classes.

Fix a compact Lie group~$G$, and suppose $BG$~is a {\it smooth\/} classifying
space.  The odd real cohomology of~$BG$ vanishes, so from~\thetag{1.7} we
conclude
  $$ \Hh^4(BG) \cong A^4_H(BG) = \{(\lambda ,\omega )\in H^4(BG)\times
     \clf^4(BG): \lambda _\RR = [\omega ]_{dR}\}. \tag{3.1} $$
Fix a connection~$A\univ$ on the universal bundle~$EG\to BG$ and suppose
$\lambda \in H^4(BG)$ is a characteristic class.  Let $\omega \univ\in
\clf^4(BG)$ be its Chern-Weil representative.  Then $(\lambda ,\omega
\univ)\in A^4_H(BG)$, and by~\thetag{3.1} this data determines a universal
{\it Chern-Simons class\/} in~$\Hh^4(BG)$.  Fix a cocycle
representative~$\alh\univ\in \Zh_H^4(BG)$.

Now suppose $P\to X$ is a principal $G$-bundle over a smooth manifold~$M$.
Let $A$~be a connection on~$P$.  A {\it classifying map\/} for~$A$ is a
$G$-equivariant map $f\:P\to EG$ such that~$f^*A\univ=A$.  It is well-known
that classifying maps exist.  Let $\fb\:M\to BG$ be the map induced from a
classifying map~$f$.  Then
  $$ \alh(A) = \fb^*\alpha \univ\in \Zh_H^4(M) \tag{3.2} $$
is the Chern-Simons cocycle of~$A$.  Note that the curvature of~$\alh(A)$ is
the Chern-Weil 4-form~$\omega (A)$ of~$A$.  As stated here $\alh(A)$~depends
on the classifying map~$f$.  Any two classifying maps are homotopic through
classifying maps, so the cohomology class of~$\alh(A)$ in~$\Hh^4(M)$ is
well-defined.  In fact, there is a more refined context in which we can work
so that $\alh(A)$~is canonically defined as a cocycle.

In 3-dimensional Chern-Simons theory~\cite{F1} one considers a family of
connections~$A$ on a compact oriented manifold~$X$ parametrized by~$T$, where
$n=\dim X\le 3$.  Then the associated classical Chern-Simons invariant is 
  $$ \int_{(X\times T)/X}\alh(A)\in \Ch_H^{4-n}(T).  $$
If $X$~is closed the result is a cocycle, so represents a differential
cohomology class.  For example, if $\dim X=3$ this cocycle is a map
$T\to\RR/\ZZ$, the usual classical Chern-Simons action.  For $\dim X<3$ we
obtain other geometric invariants. 
 
If a field theory (in arbitrary dimensions) contains a nonabelian gauge
field~$A$, we can couple a ``2-form field''~$B$ to it in a nontrivial way
using the Chern-Simons cocycle~\thetag{3.2}.  Let the underlying manifold
be~$X$.  Then we interpret the field~$B$ globally as an element ~$B\in
\Ch_H^3(X)$ which trivializes~$\alh(A)$, as explained before \theprotag{1.10}
{Example}.  Schematically, we write
  $$ d\Bh = \alh(A). \tag{3.3} $$
There are theories of~$A,B$ alone which use this coupling~\cite{BSS}, and it
appears in more complicated theories as well, as we explain next.

 \subhead Type~I $B$-field: Differential Cohomology
 \endsubhead

The coupling~\thetag{3.3} between a nonabelian gauge field~$A$ and an abelian
2-form field~$B$ occurs in type~I supergravity in 10~dimensions~\cite{CM}.  The
$B$-field occurs in {\it pure\/} supergravity, where it may be interpreted as
a cocycle in~$\Zh_H^3(X)$.  But in classical Type~I supergravity coupled to
super Yang-Mills there is also a nonabelian gauge field~$A$.  Suppose
$\alh(A)\in \Zh_H^4(X)$ is its Chern-Simons cocycle, as in~\thetag{3.2}.
Then the $B$-field is a cochain~$\Bh\in \Ch^3_H(X)$ such that $\Bh$
trivializes~$\alh(A)$: 
  $$ d\Bh = \alh(A). \tag{3.4} $$
The {\it Green-Schwarz anomaly cancellation mechanism\/} in the low energy
description of Type~I superstring theory~\cite{GS} is a modification to the
global geometric nature of~$B$ and an additional term in the action.  Here we
interpret it in differential cohomology, along the lines of the standard
story.  Namely, \thetag{3.4}~is replaced by
  $$ d\Bh=\alh(A) - \alh(g), \tag{3.5} $$
where $\alh(g)$~is the Chern-Simons cocycle of the Levi-Civita connection.
The additional term in the action has the form
  $$ \tp i\int_{X} \gh(g,A)\cdot\Bh, \tag{3.6} $$
where the curvature of the differential cocycle $\gh(g,A)\in \Zh_H^8(X)$ is
an 8-form~$P_8(g,A)$ which occurs in the anomaly computation from the
fermionic functional integrals.  (The ability to write the fermion anomaly in
this form restricts the gauge algebra to a few possibilities.  The precise
formula for~$P_8(g,A)$ may be found in~\thetag{3.38}.)  The exponential
of~\thetag{3.6} is a section of a hermitian line bundle whose curvature in a
family $X\to T$ is
  $$ \tp i\int_{X/T}P_8(g,A)\wedge\bigl[ \omega (A)-\omega (g)\bigr] . $$
(Recall that $\omega $~denotes the Chern-Weil 4-form.)  This cancels the
curvature from the fermion Pfaffian line bundles in Type~I supergravity.
Note that the existence of the trivialization~$\Bh$ in~\thetag{3.5} implies
the topological constraint
  $$ \lambda (A) = \lambda (g), \tag{3.7} $$
where $\lambda $~is the {\it integral\/} characteristic class used to define
the Chern-Simons cocycle.  Equation~\thetag{3.7} plays an important role in
heterotic string theory, for example in the cancellation of worldsheet
anomalies.  It also appears in the Type~I superstring.

In the scenario presented here the local anomaly (curvature) cancels, but
there remains a global anomaly.  Later we revise this discussion for the
Type~I superstring (and one of the heterotic strings).  We replace integral
cohomology by $KO$-theory; then the global anomaly cancels as well.
Equation~\thetag{3.7} is refined to equation~\thetag{3.46} in $KO$-theory.

 \subhead Self-dual gauge fields\footnote{We take this opportunity to point
out a conceptual mistake in~\cite{FH}.  It occurs in the paragraph following
equation~\thetag{6}, and also in the footnote which follows.  In fact, there
is no change in the quantization law of the gauge field, but rather it is the
quadratic form introduced in~\thetag{3.13} below which is needed to make
sense of the electric coupling.  See~\theprotag{3.27} {Example} for an
analogous case.  Also, there is no constraint on the Ramond-Ramond gauge
field as proposed before equation~\thetag{15} in~\cite{FH}; the factor
of~$1/2$ is implemented by the quadratic form~\thetag{3.13}.}
 \endsubhead

Often these are termed {\it chiral gauge fields\/} or {\it gauge fields with
self-dual field strength\/}; for simplicity we call them {\it self-dual gauge
fields\/}.  In {\it Lorentzian\/} field theory on~$\RR\times N$ the
self-duality condition makes sense {\it classically\/}, and it states~$F_A =
*F_A$.  This is not an equation of motion from an action principle, but
rather is an auxiliary condition imposed by hand.  The set of self-dual
solutions (up to equivalence) to the classical equations of motion is a
symplectic submanifold of the set of all solutions, and so gives a
well-defined classical system.  Note that electric and magnetic currents and
charges are equal for a self-dual gauge field: $j_E=j_B$.  In this section we
outline the additional data needed to define self-dual gauge
fields---including charge quantization---in Euclidean quantum field theory.
We assume throughout that the Riemannian manifold~$X$ is compact; if not, one
should add convergence conditions on the integrals over~$X$.  
 
There are three main examples we have in mind. 

        \example{\protag{3.8 (doubling)} {Example}}
 Here $n=\dim X$ is arbitrary.  Let $\Ah\in \ZhG^\dd(X)$ be any gauge field,
and consider now $\ddt=(\dd,n-\dd)$ and $\Aht=(\Ah,\Ah')\in \ZhG^{\ddt}(X) =
\ZhG^{\dd}(X)\times \ZhG^{n-\dd}(X)$.  The self-duality condition asserts
that $\Ah'$~is the electromagnetic dual of~$\Ah$, and it allows us to recover
the theory of~$\Ah$ from the theory of the pair $\Aht=(\Ah,\Ah')$. 
        \endexample

        \example{\protag{3.9 (integer cohomology)} {Example}}
 Here $n=\dim X=4\ell +2$ for some integer~$\ell $.  Then on a
middle-dimensional gauge field~$\Ah\in \Zh_H^{2\ell +1}(X)$ we can impose the
self-duality constraint.  Returning momentarily to the Lorentzian framework
on Minkowski spacetime~$M^n$, free theories correspond to representations of
the Poincar\'e group.  In this case we obtain an irreducible massless
representation induced from the action of the little group~$\Spin_{n-2} =
\Spin_{4\ell }$ on self-dual $2\ell $-forms. 
        \endexample

        \example{\protag{3.10 ($K$-theory)} {Example}}
 We continue \theprotag{2.10} {Example}.  Here $n=\dim X = 10$ and $X$~is
spin.  Then for vanishing $B$-field the self-duality is imposed on~$\Ah\in
\Zh_K^d(X)$.  This occurs in the low energy description of the Type~II
superstring---$\Ah$~is the {\it Ramond-Ramond gauge field\/}.
        \endexample

We list the data and constraints necessary to define a self-dual gauge
field. 

        \definition{\protag{3.11} {Definition}}
 Fix a dimension~$n$, a cohomology theory~$\Gamma $, a multi-degree
$\dd=(d_1,\dots ,d_k)$, a multi-coupling constant $e^2 = (e^2_1,\dots
,e_k^2)$, and maps~\thetag{2.14} from ~$\Gamma $ to integer cohomology~$H$ in
low degrees.  A {\it self-duality constraint\/} on the corresponding gauge
field~$\Ah$ is the additional data: 
 \medskip
 {\narrower
 \flushpar
 \rom(i\rom)\ an automorphism $\theta \:\Gh ^{\bul}\to\Gh ^{n+2-\bul}$ of a
product of $k$~copies of generalized differential cohomology so that for any
 $\Gh$-oriented fiber bundle~$X\to T$ with closed fibers of dimension~$n$,
the bilinear form
  $$ \alignedat2
      B_{X/T}\:\Zh_\Gamma ^{\dd+1}(X)&\times \Zh_\Gamma ^{\dd+1}(X)
     &&\longrightarrow \Zh_\Gamma ^{2}(T) \\
       \ah_1 \quad\; &\times  \;\quad  \ah_2\qquad &&\longmapsto
     \int_{X/T}\theta (\ah_1)\cdot \ah_2\endaligned \tag{3.12} $$
\line{is symmetric;\hfill}
  \rom(ii\rom)\ for each fiber bundle~$X\to T$ as above a quadratic map 
  $$ q_{X/T}\:\ZhG^{\dd+1}(X)\longrightarrow \ZhG^2(T) \tag{3.13} $$
which refines the bilinear map~\thetag{3.12} in the sense that there is a
natural isomorphism
  $$ q_{X/T}(\ah_1 + \ah_2) - q_{X/T}(\ah_1)- q_{X/T}(\ah_2) + q_{X/T}(0)
     \cong B_{X/T}(\ah_1,\ah_2),\qquad \ah_1,\ah_2\in
     \ZhG^{\dd+1}(X). \tag{3.14} $$
If $\bh\in \ChG^{\dd}(X)$ is a trivialization of~$\ah\in \ZhG^{\dd+1}(X)$,
then there is a canonically induced trivialization~$q_{X/T}(\bh)$
of~$q_{X/T}(\ah)$.\par}\medskip
 \flushpar
In addition, we take the specific normalizing form~$\omega _X = \bigl((\omega
_X)_1,\dots ,(\omega _X)_k \bigr)$ (see~\thetag{2.6}) defined by 
  $$ (\omega _X)_i = \sqrt{\pi e_i^2\AhG(X)}, \tag{3.15} $$
where $\AhG(X)$~is the form in~\thetag{1.15}.  Finally, the electric and
magnetic currents are constrained to satisfy 
  $$ \jh_E = \theta (\jh_B). \tag{3.16} $$
        \enddefinition

The definition requires several comments.  The quadratic form~$q$ is modeled
after~\cite{HS}, who treat \theprotag{3.9} {Example} in great detail.  They
specify a more precise set of axioms for~$q$ which we do not state explicitly
here but implicitly require.\footnote{They extend to maps
$q_{X/T}\:\ZhG^{\dd+1}(Y)\to \ZhG^i(T)$ over fiber bundles $Y\to T$ with
relative dimension~$n+2-i$ for~$i=0,1,2$.  The additional axioms involve
functoriality under base change and composition of fiber bundles.}  One
statement we will use is: The quadratic form~\thetag{3.13} is a map of
categories.  (Recall the discussion before \theprotag{1.10} {Example}.)  The
quadratic form~$q$ is determined by an analogous quadratic refinement in
$\Gamma $-cohomology, so is really a topological choice.  Typically, $q$~has
no constant term---$q_{X/T}(0)=0$---and often $q$~has no linear
term---$q_{X/T}(-\ah) = q_{X/T}(\ah)$.  However, we will encounter one
example (involving $KO$-theory) where the symmetry is of the form
$q_{X/T}(\lh-\ah)=q_{X/T}(\ah)$ for some class~$\lh$.  If the gauge field and
magnetic current live in $\zh$-twisted differential cohomology, in which case
the electric current lives in $(-\zh)$-twisted differential cohomology
(see~\thetag{2.35}), then $\theta \:\Gh^{\bul}_{\zh}\to\Gh^{n+2-\bul}_{-\zh}$
and the domains of~$B_{X/T}$ and~$q_{X/T}$ are suitably twisted.  In some
cases (e.g. \theprotag{3.22} {Example} below) the codomain of~$q_{X/T}$ does
not involve~$\Gh$ but rather a different differential theory which maps
into~$\Gh$.  In fact, we really only use the quadratic form obtained by
composing~$q_{X/T}$ with the map $\Zh_\Gamma ^2(T)\to \Zh_H^2(T)$ obtained
from~\thetag{2.14}.  The constraint~\thetag{3.16} on the currents means that
there is only one current and one charge in the theory.  This is the basic
meaning of self-duality.  Finally, we note that the quadratic refinement~$q$
of~$B$ is used twice in the self-dual theory: it determines the partition
function (see~\cite{W2}) and is used in the electric coupling term in the
action (see~\thetag{3.25} below).

        \example{\protag{3.17 (doubling)} {Example}}
 The map~$\theta $ is 
  $$ \theta (\ah,\ah') = ((-1)^{(\dd+1)(n-\dd+1)}\ah',\ah),\qquad \ah,\ah'\in
     \Gh ^{\bul}. $$
Let $\at_1=(\ah_1,\ah_1')$ and $\at_2=(\ah_2,\ah_2')$ be elements of~$\Gh
^{\dd+1}\times \Gh ^{n-\dd+1}$.  Then the bilinear form 
  $$ \at_1,\at_2\longmapsto \theta (\at_1)\cdot \at_2 = \ah\mstrut _1\ah_2' +
     \ah\mstrut _2\ah_1' \tag{3.18} $$
has a canonical quadratic refinement 
  $$ \at = (\ah,\ah')\longmapsto \ah\cdot \ah'. \tag{3.19} $$
The required~$q_{X/T}$ is obtained from this by integration. 
        \endexample

        \example{\protag{3.20 (integer cohomology)} {Example}}
 Set~ $n=4\ell +2$ and~$d=2\ell +1$.  The map~ $\theta$ is the identity.
Hopkins and Singer~\cite{HS}, following Browder~\cite{Br}, explain that on
the category of compact oriented manifolds with ``Wu structure'' there is a
functorial quadratic refinement defined.  In the familiar case~$n=2$ of a
self-dual scalar field, a Wu structure is simply a spin structure.  The
normalization~\thetag{3.15} corresponds to the ``free fermion radius''.
Namely, with~$\omega _X=\tp$ so that the gauge field~$\phi $ has
periodicity~$\tp$, the kinetic action is
  $$ S_{\text{kin}}(\phi ) = \frac{1}{8\pi }\int_{X}d\phi \wedge *d\phi
     . \tag{3.21} $$
        \endexample

        \example{\protag{3.22 ($K$-theory)} {Example}}
 Recall $n=10$ and $X$~is spin.  Then
  $$ \aligned
      \theta \:\Kh^q(X)&\longrightarrow \Kh^{12-q}(X) \\
      \ah&\longmapsto u^{6-q}\bar{\ah}\endaligned \tag{3.23} $$
where $u\in K^2(pt)$ is the inverse Bott element.  If $\ah\in \Kh^0(X)$~is
represented by a complex vector bundle $E\to X$ with connection, then
$\bar{\ah}$~is represented by the complex conjugate vector bundle
$\overline{E}\to X$ with the conjugate connection.  The quadratic refinement
is defined by Witten in\footnote{Witten actually defines a related quadratic
form, but the basic idea is the same.}~\cite{W4}, and it uses the fact that
we integrate over Riemannian {\it spin\/} fiber bundles $X\to T$.  Namely,
for~$\ah\in \Zh_K^\bul(X)$ the element~$\theta (\ah)\cdot \ah$ has a
canonical lift to~$\Zh_{K@!@!@!Sp}^{12}(X)$, and topologically
$q_{X/T}(a)$~is the pushforward of this lift in $\KSph$-theory.
        \endexample

In the quantum theory the Euclidean partition function and correlation
functions of self-dual fields are not defined by the usual functional
integral, but rather a special procedure is needed which accounts for the
self-duality constraints.  The reader should keep in mind that the
interpretation of this action is not that of the usual functional integral.
 
We work in the situation of \theprotag{3.11} {Definition}.  Let $\jh\in
\ZhG^{\dd+1}(X)$ and set $\jh_B=\jh$, $\jh_E=\theta (\jh)$.  The gauge
field~$\Ah\in \ChG^\dd(X)$ is a trivialization of the refined magnetic
current~$\jh$.  The kinetic term in the action is unchanged
from~\thetag{2.11}, but we do use the normalization~\thetag{3.15}.  For a
gauge field quantized by integer cohomology with~$\omega _X=\tp$ we find 
  $$ S_{\text{kin}}(A) = \frac{1}{8\pi }\int_{X}F_A\wedge *F_A,  $$
generalizing~\thetag{3.21}.  In the presence of the self-duality constraint
the electric coupling term is half\footnote{Some justification for the
factor~1/2 comes by considering the classical Lorentzian field theory;
see~\cite{FH,(6)}.} the usual term~\thetag{2.16}.  Recall first the
anomaly~\thetag{2.30}; in the self-dual case it is the exponential of $\tp
i$~times 
  $$ \text{``}-\frac 12\int_{X/T}\theta (\jh)\cdot\jh = -\frac 12\;
     B_{X/T}(\jh,\jh)\text{''}  $$
where we use the degree~2 map in~\thetag{2.14} to land in~$\Hh^2(T)$.  Taking
half the integral is precisely what the quadratic form~$q_{X/T}$ does, so the
self-dual anomaly is
  $$ \exp\Bigl( -\tp i\;q_{X/T}(\jh)\Bigr). \tag{3.24} $$
The electric coupling term is formally the exponential of $\tp i$~ times 
  $$ \text{``}-\frac 12\int_{X/T}\theta (\jh) \cdot \Fh_A= -\frac 12\;
     B_{X/T}(\jh,\Fh_A),\text{''}  $$
which is 
  $$ \exp\Bigl( -\tp i\;q_{X/T}(\Fh_A)\Bigr). \tag{3.25} $$
This is a trivialization of the circle bundle~\thetag{3.24}.  The entire
exponentiated action is
  $$ e^{-S}(\Ah ) = \exp\bigl(-\frac 1{8\pi }\int_{X}F_A\wedge *F_A
     \bigr)\,\exp\bigl(-\tp i\;q_{X/T}(\Fh_A) \bigr). \tag{3.26} $$

        \example{\protag{3.27 (doubling)} {Example}}
 It is instructive to work out the quadratic map~\thetag{3.13} in the case of
a pair of $\TT$-valued scalar fields, a special case of \theprotag{3.8}
{Example} and \theprotag{3.17} {Example}.  Thus $\Gamma =H$~is integer
cohomology and the degree is~$\dd=(1,1)$.  Let the dimension be~$n=2$.
Assume $X\to T$ has fibers which are closed oriented surfaces.  We use
formulas~\thetag{3.18} and~\thetag{3.19} for the bilinear form and its
quadratic refinement.  Consider first~$\jh=0$ so that the gauge field
$\Ah=(e^{i\phi },e^{i\phi '})$ is a pair of maps~$X\to\TT$.  Then the
anomaly~\thetag{3.24} vanishes so the electric coupling~\thetag{3.25} is an
ordinary function~$T\to\TT$.  The square of this function is identically~1 as
it is the exponential of~$\tp i$ times~$-B_{X/T}(\jh,\Fh_A)$.  Thus
\thetag{3.25}~is a locally constant function with value~$\pm 1$.  In fact, it
is the exponential of $\tp i$~times
  $$ -\frac 12\int_{X/T}\frac{d\phi }{\tp}\wedge \frac{d\phi
     '}{\tp}. \tag{3.28} $$
Note that $\frac{d\phi }{\tp}\wedge \frac{d\phi '}{\tp}$ represents an
integral cohomology class on~$X$, the characteristic class of the
product~$e^{i\phi }\cdot e^{i\phi '}$ in~$\Hh^2(X)$.
 
Now consider $\jh=(0,\kh')$ where $\kh'\in \Hh^2(X)$ is represented by a
circle bundle $P'\to X$ with connection.  Then the gauge field
is~$\Ah=(e^{i\phi },f')$, where $e^{i\phi }\:X\to\TT$ and $f'$~is a section
of~$P'$.  The exponential of $\tp i$~times
$B_{X/T}(\jh,\Fh_A)=\int_{X/T}e^{i\phi }\cdot \kh'$ is again an ordinary
function~$T\to\TT$, the anomaly $\exp\bigl(-\tp i\;q_{X/T}(\jh) \bigr)$
vanishes, and $\exp\bigl(-\tp i\;q_{X/T}(\Fh_A) \bigr)$ is an ordinary
function---a square root of $\exp\bigl(-\tp i\int_{X/T}e^{i\phi }\cdot \kh'
\bigr)$.  Let $\theta \in \Omega ^1(X)$ be the covariant derivative of~$f'$;
then $d\theta $~is the curvature of~$P'$.  A computation in differential
cohomology yields
  $$ \exp\bigl(-\tp i\;q_{X/T}(\Fh_A) \bigr) = \exp\bigl(-\frac{\tp i}{2}
     \int_{X/T} \frac{d\phi }{\tp}\wedge
     \frac{\theta}{\tp}\bigr), \tag{3.29} $$
a generalization of~\thetag{3.28}.  As a special singular case let $P'$~be
flat outside a ``divisor'' of degree~0 on each component of~$X$.  For
example, let $P'$~have curvature~$\delta _p-\delta _q$ where $p$~ and $q$~are
sections of~$X\to T$ whose images lie in the same component and $\delta
_p$~and $\delta _q$~are distributional 2-forms supported on these images.
Then (compare~\thetag{2.24})
  $$ \exp\bigl(-\tp i\;B_{X/T}(\Fh_A,\jh) \bigr) = \frac{e^{i\phi
     (q)}}{e^{i\phi (p)}}.  $$
\thetag{3.29}~is a square root of this function, but as $f'$~is discontinuous
at~$p,q$ it is not easy to describe geometrically.
        \endexample

 \subhead Type~II Ramond-Ramond fields ($B=0$)
 \endsubhead

As described in \theprotag{2.10} {Example}, in the low energy description of
the Type~II superstring $X$~is a spin 10-manifold.  If the $B$-field
vanishes, then the Ramond-Ramond gauge field~$\Ah$ lives in~$\Ch_K^\bul(X)$
with $\bul=0$ for Type~IIA and $\bul=-1$ for Type~IIB.  Furthermore, the
gauge field is self-dual and the extra data (see \theprotag{3.11}
{Definition}) needed to describe it is specified in \theprotag{3.22}
{Example}.  Our goal here is to make everything a bit more explicit for the
current~$\jh$ induced from a submanifold $i\:W\hookrightarrow X$, the
worldvolume of a {\it D-brane\/}.
 
Suppose $W$~has codimension~$r$ in~$X$; $r$~is odd in Type~IIA and even in
Type~IIB.  We assume given $\qh\in \Zh^0_{K,\th }(W)$, analogous
to~\thetag{2.19} and~\thetag{2.26}, but where $\th $~is a twisting of
differential $K$-theory.  (Twistings are discussed at the end of~\S{1} and in
this context at the end of~\S{2}.)  Define the magnetic current as
  $$ \jh = u^{-\left[ \frac p2 \right] }\,i_*\qh. \tag{3.30} $$
Concretely (see \theprotag{1.13} {Example}) $\qh$~is usually described as a
complex vector bundle with connection $Q\to W$, the ``Chan-Paton vector
bundle''.  For the pushforward to be well-defined we have
equation~\thetag{2.34} for the twistings.  Recall from~\thetag{1.17} that the
twisting class in $\Kh$-theory of a real vector bundle~$V$ is the
pair~$\bigl(w_1(V),\wh_2(V) \bigr)$.  Now since $X$~is assumed spin, its
twisting class vanishes.  Hence \thetag{2.34}~asserts that $\qh$~is a twisted
cocycle on~$W$, the twisting being~$\bigl(w_1(\nu ),\wh_2(\nu ) \bigr)$,
where $\nu \to W$ is the normal bundle to~$W$ in~$X$.  Usually one assumes
$W$~is oriented, in which case the twisting is by~$\wh_2(\nu )$.  For
example, a (locally) rank one element~$\qh$ is not represented by a complex
line bundle over~$W$ with connection, but rather by a \spinc~connection
on~$\nu $.  This was derived from perturbative string theory in~\cite{FW}.
 
Next we work out the electric coupling term~\thetag{3.25} and the associated
anomaly.  First, the electric current is 
  $$ \theta (\jh) = u^{e - \left[ \frac p2 \right] }\,i_*\qhb,  $$
where $e=5$ in Type~IIA and $e=6$ in Type~IIB.  The electric coupling term is
a section of a hermitian line bundle with connection over a parameter
space~$T$; the line bundle represents the anomaly.  We compute the Chern
class of this line bundle from~\thetag{3.24} as
  $$ -q_{X/T}(\jh) = -\int_{X/T}u^{6-p}\,i_*\qhb\cdot i_*\qh,  $$
where the integral is in~$\KSph$.  We can write this as an integral
over~$W/T$
  $$ \split
      -q_{X/T}(\jh) &= -\int_{W/T}u^{6-p}\,\qhb\cdot i^*i_*\qh \\
      &= -\int_{W/T}u^{6-p}\,\qhb\cdot \qh\cdot \Euler_{\Kh}(\nu ),\endsplit
      $$
but the integrals can no longer be interpreted in~$\KSph$.  The last section
in~\cite{FH} explains how to interpret this computation depending on the
dimension of~$W$, and also gives a formula for the $\Kh$-theory Euler class.
(The subtle point is while for $r$~odd the Euler class vanishes in
topological $K$-theory, it is an element of order two in differential
$\Kh$-theory.)  As explained there, this anomaly cancels the anomaly from the
fermions on~$W$.  The electric coupling~\thetag{3.25} may be written formally
as
  $$ \tp i\;q_{X/T}(\Fh_A)\;\; \text{``}=\text{''} \;\;\frac {\tp
     i}2\int_{X/T}i_*\qhb\cdot\Fh_A  \;=\; \frac{\tp
     i}{2}\int_{W/T}\qhb\cdot i^*\Fh_A.\tag{3.31} $$
The integrals are in $\Kh$-theory and the factor of~$1/2$ is because of the
quadratic form.  The electric coupling appears in this form
in~\cite{FH,\thetag{15}}.  To convert to a formula with differential forms,
we assume that $\qh$~is defined by a complex vector bundle $Q\to W$ with
connection.  Suppose also that $W$~is \spinc\ and the curvature of the
\spinc\ connection is $-\tp i\eta \in \Omega ^2(W)$.  Finally, suppose the
Ramond-Ramond field is determined by a differential
form~$A/\bigl(\,\tp\sqrt{\Ahat(X/T)}\,\bigr)$ with~$dA=F_A$, at least
over~$W$ (see~\thetag{1.8}).  Then \thetag{3.31}~reduces to
  $$ \tp i\;q_{X/T}(\Fh_A) = \pi i\int_{W/T}\Ahat(W/T)\wedge e^{\eta
     /2}\wedge \ch(\overline{Q}) \wedge i^*\left(
     \frac{A}{\tp\sqrt{\Ahat(X/T)}}\right). \tag{3.32} $$
The formula appears in roughly this form in the physics literature
(e.g.~\cite{MM}, \cite{CY}).  Notice that we ignore the magnetic current in
writing this expression.  We have already given a precise definition of the
electric coupling; \thetag{3.32}~is included to make contact with the
literature.

 \subhead Type II Ramond-Ramond fields ($B\not= 0$)
 \endsubhead

The Type~II $B$-field is a cocycle~$\Bh\in \Zh_H^3(X)$.  In other words, it
is a ``usual'' 2-form gauge field quantized by integer cohomology.  As
explained at the end of~\S{1} it determines a twisted
version~$\Kh^{\bul+\Bh}(X)$ of differential $K$-theory, and the Ramond-Ramond
fields are cochains in this twisted theory.  The Ramond-Ramond charges take
values in the twisted $K$-group~$K ^{\bul+\zeta}(X)$, where $\zeta \in
H^3(X)$ is the characteristic class of the $B$-field~$\Bh$.
 
The previous discussion may be reconsidered with this twist.  The
automorphism~$\theta $ has the same formula~\thetag{3.23} as before, but it
reverses the twisting: 
  $$ \theta \:\Kh^{q+\Bh}(X)\longrightarrow \Kh^{12-q-\Bh}(X)
     .  $$
By~\thetag{2.36} the twisting~$\th$ of the Chan-Paton bundle~$\qh\in
\Zh_{K}^{0+\th}(W)$ satisfies
  $$ \th = \wh_2(\nu ) + i^*\Bh. \tag{3.33} $$
This equation was deduced from perturbative open string theory in~\cite{FW};
it is one of many pieces of evidence that Ramond-Ramond charge lives in
$K$-theory.  Equation~\thetag{3.33} is a nontrivial constraint on D-branes.
Traditionally one thinks of the Chan-Paton vector bundle on a ``single''
D-brane as having rank one.  The concept of rank does not make sense in every
twisted $K$-theory, and the only reasonable interpretation is that a rank one
element is a cochain in~$\Ch^2_H(W)$ which trivializes~$\th\in \Zh_H^3(W)$.
Such trivializations exist if and only if the topological class of the
twisting vanishes:
  $$ W_3(\nu ) + i^*\lambda (\Bh) =0. \tag{3.34} $$
This constraint on a single D-brane was derived from different points of view
in~\cite{W5}, \cite{W3}, and~\cite{FW}.  If the class $W_3(\nu ) + i^*\lambda
(\Bh)$ is torsion of order~$N$, then again it makes sense to talk about
twisted $K$-theory elements of finite rank, but the rank is constrained to be
a multiple of~$N$.  Elements of virtual rank zero exist for any twisting; one
might instead formally consider them to have infinite rank~\cite{W1}.
  
Explicit formulas for twisted $\Kh$-theory are difficult to write in general,
but can be written if we assume~$\Bh$ to be defined by a global real
2-form~$B\in \Omega ^2(X)$ (cf.~\thetag{1.8}).  Note that the characteristic
class~$\zeta $ of this~$\Bh$ vanishes and the curvature is~$dB$.  The
form~$B$ induces a map 
  $$ \psi \:\Zh^2_H(X) \longrightarrow \Ch^2_H(X) \tag{3.35} $$
whose image consists of trivializations of~$\Bh$.  In the model of
\theprotag{1.10} {Example} the triple~$(0,B,dB)$ represents~$\Bh$ and the
map~\thetag{3.35} is 
  $$ \psi \:(c,h,\omega ) \longmapsto (c,h,\omega +B).  $$
In our current notation, if $\Ah\in \Zh^2_H(X)$ has field strength~$F_A$,
then the field strength (covariant derivative) of~$\psi (A)$ is~$F_A+B$.
Note that trivializations of~$\Bh$ are ``rank one'' cocycles
for~$\Kh^{\bul+\Bh}(X)$.
 
We can apply these remarks to construct~$\qh\in \Zh^{0+i^*(\Bh)}_{K}(W)$ in
case $W$~is spin.  Thus we suppose $\Ah\in \Zh_H^2(W)$ is an ordinary 1-form
gauge field on~$W$ and set $\qh=\psi (\Ah)$.  In explicit formulas
like~\thetag{3.32} the field strength~$F_A$ is replaced by $F_A+B$, for
example in the factor~$\ch(\overline{Q})$.  Also, these remarks explain a
puzzle~\cite{BDS} about Ramond-Ramond charge which was resolved in~\cite{T},
\cite{AMM}.  Namely, since the characteristic class of~$\Bh$ vanishes the
Ramond-Ramond charges take values in ordinary $K$-theory: $\Bh$-twisted
topological $K$-theory is not twisted.  The remarks above tell that the
$\Bh$-twisted $K$-theory class of $\jh=u^{-\left[ \frac p2 \right] }\,i_*\qh
\in \Zh^{\bul+\Bh}_{K}(X)$ (see~\thetag{3.30}) is the ordinary $K$-theory class
of~$i_*\Ah$.  So in explicit formulas for the Ramond-Ramond charge---as in
the papers just cited---one finds the field strength~$F_A$, not~$F_A+B$.

 \subhead Type I $B$-field: Differential $KO$-Theory
 \endsubhead

We have already discussed the $B$-field in Type~I superstring theory and the
Green-Schwarz local anomaly cancellation from the point of view of integral
cohomology.  But according to~\cite{W3} the charges in Type~I superstring
theory lie in $KO$-theory.  Hence we expect the $B$-field to be interpreted
in differential $KO$-theory.  In fact, this 2-form field is related to the
Ramond-Ramond field in Type~IIB, and this also leads us to expect that the
corresponding charge is quantized in terms of some form of $K$-theory.  Since
the Ramond-Ramond fields are self-dual, we expect that in the differential
$KO$ formulation the Type~I $B$-field is also self-dual.  Finally, the
Atiyah-Singer index theorem computes the fermion anomaly as an integral in
differential $KO$-theory.  For this anomaly to cancel against local and
global anomalies involving bosonic gauge fields, we expect the gauge fields
to be cochains in differential $KO$-theory.  We develop these ideas in this
section.  Proofs of some mathematical assertions made in this discussion are
deferred to Appendix~B, written jointly with M. Hopkins.
 
Let $X\to T$ be a Riemannian spin fiber bundle with fibers closed
10-manifolds.  Recall this means that there is a Riemannian metric on the
relative tangent bundle~$T(X/T)$ and a distribution of horizontal planes
on~$X$, as well as a spin structure on~$T(X/T)$.  In Type~I superstring
theory there is a real rank~32 vector bundle $E\to X$ with connection~$A$.
The fermion anomaly has three contributions: a chiral spinor field with
values in~${\tsize\bigwedge} ^2E$, the adjoint bundle to~$E$; a chiral
Rarita-Schwinger field, which is a chiral spinor field coupled to~$T(X/T)-1$;
and a chiral spinor field of the opposite chirality.  (The trivial bundle~$1$
is subtracted from the relative tangent bundle to obtain the pure spin-3/2
field.)  The fermion anomaly is a complex line bundle $\scrL\to T$ with
connection, and it is computed by a geometric form of the index
theorem~\cite{BF}.  We express\footnote{As mentioned at the end of
\theprotag{1.13} {Example}, the rigorous derivation of formulas
like~\thetag{3.36} is part of an ongoing project with M. Hopkins and
I. Singer.} the answer in differential $KO$-theory:
  $$ \scrL = \pfaff\int_{X/T}\,{\tsize\bigwedge} ^2\Eh + \Th(X/T) -
     2. \tag{3.36} $$
Here $\Eh\in \Zh^0_{KO}(X)$ is the cocycle corresponding to the real vector
bundle with connection~$E$; similarly, $\Th(X/T)\in \Zh^0_{KO}(X)$; the
integral is a map $\int_{X/T}\:\Zh^0_{KO}(X)\to \Zh^{-10}_{KO}(T)$; and
$\pfaff\:\Zh^{-10}_{KO}(T)\to \Zh^2_{KO}(T)$ is the pfaffian line bundle.
The standard formula in the physics literature (\cite{GSW,\S13.5}, for
example) is for the curvature of~$\scrL$, which we write as
  $$ \curv\scrL=\tp i\int_{X/T} \frac 12\, P_8(g,A) 
     \wedge \bigl[p_1(g) - \ch_2(A) \bigr], \tag{3.37} $$
where 
  $$ P_8(g,A) = -\ch_4(A) +\frac{1}{48}p_1(g)\,\ch_2(A) -
     \frac{1}{64}p_1(g)^2 + \frac{1}{48}p_2(g). \tag{3.38} $$
The integrand in~\thetag{3.37} is a rational combination of Chern-Weil
differential forms for the Pontrjagin classes of~$T(X/T)$ and the Chern
character classes of the complexification of~$E$.  The extra factor of~$1/2$
is due to the fact that $\scrL$~is the {\it pfaffian\/} line bundle, a square
root of the determinant line bundle.  The factorization of the integrand
in~\thetag{3.37} is a crucial ingredient in the standard story.  Usually
$P_8(g,A)$~is expressed in terms of characteristic forms
of~${\tsize\bigwedge} ^2E$ rather than~$E$; in that case there is a term
$\bigl[ \ch_2({\tsize\bigwedge} ^2A)\bigr]^2$.  The fact that
\thetag{3.38}~is affine linear in~$A$ is important to our argument.  Set
  $$ \TTh = \Th(X/T) + 22.  $$ 
When $\Eh=\TTh$ the curvature of~$\scrL$ vanishes, as the first factor in the
integrand does.  We claim, and provide a proof in \theprotag{B.1}
{Proposition}, that $\scrL$~itself is trivial for~$\Eh=\TTh$, and so in
general we can rewrite the formula~\thetag{3.36} for the fermion
anomaly~$\scrL$ as
  $$ \scrL \cong \pfaff\int_{X/T} {\tsize\bigwedge} ^2\Eh - {\tsize\bigwedge}
     ^2\TTh. \tag{3.39} $$

We turn now to the $B$-field, a local 2-form field whose global description
we now make precise.  The charges associated to this gauge field lie
in\footnote{As in Type~IIB, charges lie in the {\it augmentation
ideal\/}~$\epsilon \inv (0)$, where $\epsilon \:KO^0(X)\to H^0(X)$ is the
augmentation.}~$KO^0(X)$, so the gauge field~$\Bh$ is at first glance a
cocycle for~$(\KOh)\inv (X)$.  In fact, there is a background magnetic
current, which we have already seen in a different scenario in~\thetag{3.5}.
We now give the self-duality data of \theprotag{3.11} {Definition}.  The
cohomology theory underlying this example is the $(\ZZ/2\ZZ\times
\ZZ)$-graded theory~$KOSp$ which was described in \theprotag{1.4} {Example}.
The automorphism~$\theta $ in the degree we need is
  $$ \aligned
      \theta \:(\KOh)^0(X) &\longrightarrow (\KSph)^{12}(X) \\ 
      \ah &\longmapsto u^6\ah.\endaligned  $$
(Recall that $u^6$~is quaternionic.)  For the quadratic
refinement~\thetag{3.13} of the bilinear form~\thetag{3.12} we set 
  $$ \aligned
      q_{X/T}\: \Zh^0_{KO}(X) &\longrightarrow \Zh^2_{KSp}(T) \\ 
      \ah &\longmapsto \int_{X/T} u^6\,\lambda ^2(\ah),\endaligned \tag{3.40}
     $$
where $\lambda ^2\:\Zh_{KO}^0(X)\to\Zh_{KO}^0(X)$ is the second exterior
power operation and the integral is a map $\int_{X/T}\:\Zh_{KSp}^{12}(X)\to
\Zh_{KSp}^2(T)$.  If $\ah$~is the $\KOh$-cocycle of a real vector bundle with
connection~$E$, then $\lambda ^2(\ah)$ is the $\KOh$-cocycle
of~${\tsize\bigwedge} ^2E$ with the induced connection.  The normalizing form
is
  $$ \omega _X = \tp\sqrt{\Ahat(X)},  $$
and the coupling constant is~$e^2=4\pi $. 
 
The quadratic form~$q_{X/T}$ satisfies $q_{X/T}(0)=0$, but it is not
symmetric about the origin.  In Appendix~B we define a $\mu(X)\in KO^0(X)$
canonically associated to spin 10-manifolds~$X$.  The class $2\mu (X)\in
KO^0(X)$ is a $KO$-theoretic analog of the Wu class in cohomology.  If
$X$~has a Riemannian structure then there is a canonical lift of~$2@,\mu (X)$
to differential $KO$~theory (and a canonical cocycle representative), but
there may be many lifts of~$\mu (X)$ to a differential $KO$~class $\mh(X)$.
(An analogy: Square roots of the canonical bundle of a Riemann surface exist,
but none is canonically picked out.)  We term a choice of~$\mh(X)$ a {\it
$\mh$-structure\/}.  Furthermore, we only consider families $X\to T$ in which
an appropriate class $\mh(X/T)$~is defined.  These ideas are developed in
Appendix~B, where the following facts are part definition, part proposition:
 \roster
 \item "$\bul$" The quadratic form~$\pfaff q_{X/T}$ has a symmetry: 
  $$ \pfaff q_{X/T}(2\mh-\ah) \cong \pfaff q_{X/T}(\ah),\qquad \ah\in
     \Zh^0_{KO}(X). \tag{3.41} $$
In other words, $\mh$~is a center for~$\pfaff q_{X/T}$.
 \item "$\bul$" Set 
  $$ \eh = \mh - \TTh.  $$
Then $\eh$~restricts to zero on the 7-skeleton of~$X$.  For example, the
Chern character of~$\eh$ contains only forms of degree~$\ge8$.  There is an
explicit formula for~$\eh$ after inverting~2:
  $$ \eh = \frac {1}{64}\bigl( 7\lambda ^2(\Th) - 5\Sym^2(\Th) + 4\Th - 80
     \bigr) + \cdots , \tag{3.42} $$
where $\Th = \Th(X/T)$ is the relative tangent bundle. 
 \item "$\bul$" We have 
  $$ \pfaff q_{X/T}(\mh) \cong \pfaff q_{X/T}(\TTh) \cong
     \pfaff\int_{X/T}{\tsize\bigwedge} ^2\TTh. \tag{3.43} $$
 \endroster
 \flushpar 
 From the formula~\thetag{3.42} for~$\mh$ we compute the Chern character 
  $$ \ch(\mh) = 32 + p_1(g)\,u^{-2} + \left[ \frac{1}{192}p_1(g)^2 +
     \frac{1}{48}p_2(g) \right]u^{-4} + \cdots .   $$
Also, from~\thetag{3.14} and~\thetag{3.41} we deduce
  $$ \pfaff \,2q_{X/T}(\ah) \cong \pfaff B_{X/T}(\ah,\ah-2\mh),\qquad \ah\in
     \Zh^0_{KO}(X). \tag{3.44} $$
 
If $X^{10}\cong \RR^3\times Y^7$ then according to the second point above
there is a canonical~$\mh$-structure $\mh(X)= \TTh(X)$.  Furthermore, in this
case the characteristic class $[\TTh]\in KO^0(X)$ is determined by:
$\rank(\TTh)=32$; $w_1(\TTh) = w_2(\TTh)= 0$; and $\lambda (\TTh)\in
H^4(X)$, where $\lambda $~is the canonical~$\frac 12p_1$ for spin bundles.

Our postulate for the 2-form field in Type~I on a Riemannian spin
10-manifold~$X$ is:
  $$ \text{$\Bh$~is a nonflat isomorphism $\mh(X)\longrightarrow
     \Eh$}. \tag{3.45} $$
The notion of nonflat trivialization was explained in the paragraph
following~\thetag{1.9}; it is the same notion which underlies
equation~\thetag{3.5}.  A nonflat isomorphism is similar.  One version is
this: $\mh(X)$~and $\Eh$~are elements of a category and $\Bh$~is a morphism
between~$\mh(X)$ and~$\Eh-H$ for some global differential form~$H$
(see~\thetag{1.8}).  Another version uses the bigraded theory referred to in
the footnotes of~\S{1}.  The form~$H$ has components in degrees~3 and~7 on a
10-manifold.  Assertion~\thetag{3.45} means that there is a background
magnetic current equal to~$\Eh-\mh(X)$.  A necessary and sufficient condition
for~$\Bh$ to exist is that
  $$ \bigl[E\bigr] = \mu (X)\qquad \text{in $KO^0 (X)$}. \tag{3.46} $$
From the remark in the previous paragraph, we see that \thetag{3.46}~ is
equivalent to the standard condition~\thetag{3.7} if $X = \RR^3 \times Y^7$;
in many cases it is a stronger condition.  There is also a background
electric current, manifested through an electric coupling term of the
form~\thetag{3.25}.  We write it for a family~$X\to T$ of Riemannian spin
10-manifolds for which $\mh=\mh(X/T)$ exists.  Recall that in~\thetag{3.25}
the map $(\KSph)^2\to\Hh^2$ is omitted from the notation and the factor~$\tp
i$ is part of the identification of~$\Hh^2$ with connections on circle
bundles.  Now since $\Bh$ is a nonflat isomorphism $\mh\to\Eh$, by applying
the quadratic form, which is a map of categories, we obtain a nonflat
isomorphism $q_{X/T}(\Bh)\:q_{X/T}(\mh)\to q_{X/T}(\Eh)$.  Thus the
background electric coupling term---the {\it inverse\/}\footnote{due to the
minus sign in~\thetag{3.25}.} pfaffian applied to~$q_{X/T}(\Bh)$---is a
nonflat trivialization of
  $$ \pfaff\,-\bigl\{q_{X/T}(\Eh) - q_{X/T}(\mh)\bigl\}. \tag{3.47} $$
In other words, the anomaly in the background electric coupling term
is~\thetag{3.47}.  Using~\thetag{3.40}, \thetag{3.43}, and Bott periodicity
in~$KO$-theory we see that \thetag{3.47}~precisely cancels the
anomaly~\thetag{3.39} from the fermions.
 
There is also a version of Type~I theories in which $E$~is a projective
bundle with a nontrivial cocycle $w\in H^2(X;\zt)$.\footnote{It is usually
asserted that the gauge group of Type~I is~$\Spin_{32}/(\zt)$; the
cocycle~$w$ is the obstruction to lifting the associated
$\Spin_{32}/(\zt\times \zt)$~bundle to an $SO_{32}$~bundle.}  This means that
$\Eh$~lies in a twisted version of~$\KOh$, namely $\Eh\in \Zh^{0+w}_{KO}(X)$.
Note that $w$~is a torsion version of the 2-form field (called~`$B$') in
Type~II.  It seems that arguments parallel to those in Appendix~B yield a
``twisted'' $KO$~class $\mu_w(X)\in KO^{0+w}(X)$ associated to a spin
10-manifold, and so a parallel discussion with $\mh$~replaced by~$\mh_w$, but
we do not discuss such twisted classes in this paper.

Our motivation for~\thetag{3.45} is not simply the anomaly cancellation.
After all, because of~\thetag{3.43} we could substitute~$\TTh$ for~$\mh(X)$
in~\thetag{3.45} and still cancel the anomaly.  An additional motivation for
the choice of~$\mh(X)$ is that it is the choice which renders the magnetic
current equal to the electric current, which we require for a self-dual
field.  Another motivation is the presumed twisted analog of $\mh$ just
described; there is no such twisted analog of~$\TTh$, for example.

To make contact with the usual presentation of the {\it local\/} anomaly
cancellation, we now relate the electric coupling to the standard formula for
the Green-Schwarz term.  We write~$\Bh$ as a differential
form~$\bigl({B_2u^{-2} + B_6u^{-4}}\bigr)/
\bigl({\tp\sqrt{\Ahat(X/T)}}\bigr)$ relative to a fixed trivialization of the
background magnetic current~$\Eh-\mh$.  The differential of the covariant
derivative $\bigl({H_3u^{-2} + H_7u^{-4}}
\bigr)/\bigl({\tp\sqrt{\Ahat(X/T)}}\bigr)$ is the Chern character of the
background magnetic current, and so
  $$ \spreadlines{6pt}\aligned
      d\left( \frac{H_3u^{-2}+H_7u^{-4}}{\tp} \right) &= \sqrt{\Ahat(X/T)}
     \;\ch(\Eh-\mh) \\
      &= \Bigl[ \ch_2(A) - p_1(g)\Bigr] u^{-2}\\
       &\qquad \qquad +\[\ch_4(A) -
     \frac{1}{48}\,p_1(g)ch_2(A) + \frac{1}{64}\,p_1(g)^2 -
     \frac{1}{48}\,p_2(g)\]u^{-4}.\endaligned \tag{3.48} $$
The fixed global trivialization of the background magnetic current gives a
fixed trivialization of the anomaly~\thetag{3.47}, relative to which the
electric coupling term may be written as an integral of differential forms.
Set~$\ah = \Eh - \mh$.  Then
  $$ \split
      q_{X/T}(\Eh) - q_{X/T}(\mh) &= q_{X/T}(\mh+\ah) - q_{X/T}(\mh) \\ 
      &= q_{X/T}(\ah) + B_{X/T}(\mh,\ah),\endsplit  $$
so combining with~\thetag{3.44} we find 
  $$ \pfaff \, 2\bigl[ q_{X/T}(\Eh) - q_{X/T}(\mh) \bigr] = \pfaff \,
     B_{X/T}(\ah,\ah). \tag{3.49} $$
Thus the electric coupling in the unexponentiated action is
  $$ \spreadlines{12pt} \multline
      (\tp i)\,\frac 12 u^6\int_{X/T} \[ \Ahat(X/T)\wedge \ch(\ah)\wedge
     \(\frac{B_2u^{-2} + B_6u^{-4}}{\tp\sqrt{\Ahat(X/T)}}\) \]_{(10)}\\
      = (\tp i)\,\frac 12u^6\int_{X/T} \Biggl[\sqrt{\Ahat(X/T)} \wedge
     \[\ch(\Eh - \mh) \] \wedge \(\frac{B_2 u^{-2}+
     B_6u^{-4}}{\tp}\)\Biggr]_{(10)}. \endmultline\tag{3.50} $$
The factor of~$1/2$ in~\thetag{3.50} is due to the factor of~2
in~\thetag{3.49}; the $u^6$~is in~\thetag{3.40}.  We expand
$\sqrt{\Ahat(X/T)} \wedge \ch(\Eh-\mh)$ using~\thetag{3.48}.

Now $(B_2,B_6)$ is a self-dual pair of gauge fields (with no Dirac
quantization condition).  The self-dual Euclidean action has the form
  $$ \frac 12 \,\kinetic(B_2) + \frac 12\,\kinetic(B_6) + \frac{ \tp
     i}{2}\int_{}\;\frac{j_E}{\tp} \wedge \frac{B_2}{\tp}+ \frac{\tp
     i}{2}\int_{}\;\frac{j_B}{\tp} \wedge \frac{B_6}{\tp}  $$
and the magnetic currents are determined by 
  $$ \aligned
      dH_3 &= j_B, \\ 
      dH_7 &= j_E.\endaligned  $$
The ``$\frac 12$~kinetic'' terms are one-half the value for non-self-dual
gauge fields.  If we eliminate~$B_6$ and write the same system in terms of a
single gauge field~$\tilde B_2$, then the corresponding Euclidean action is
  $$ \kinetic(\tilde{B}_2) + \tp i\int_{}\;\frac{j_E}{\tp}\wedge\frac{\tilde
     B_2}{\tp} \tag{3.51} $$
with magnetic current determined by
  $$ d\tilde H_3 = j_B. \tag{3.52} $$
From~\thetag{3.48} and~\thetag{3.50} we read off 
  $$ \aligned
      j_B &= \tp\Bigl[\ch_2(A) - p_1(g)\Bigr], \\ 
      j_E &= \tp\Bigl[\ch_4(A) -
     \frac{1}{48}\,p_1(g)\ch_2(A) + \frac{1}{64}\,p_1(g)^2 -
     \frac{1}{48}\,p_2(g)\Bigr],\endaligned  $$
Thus \thetag{3.52}~yields
  $$ d\(\frac{\tilde H_3}{\tp}\) = \ch_2(A) - p_1(g),  $$
in agreement with~\thetag{3.5}.  The electric coupling in~\thetag{3.51} is
the Green-Schwarz term
  $$ -\tp i\int_{X/T}P_8(g,A)\wedge\frac{\tilde B_2}{\tp} , $$
where $P_8(g,A)$~is given in~\thetag{3.38}.  Therefore, our electric coupling
term is indeed a refinement of the standard Green-Schwarz term.
 
Finally, we consider the anomaly for D1- and D5-branes in Type~I; the
discussion is similar to the treatment~\cite{FH} of D-brane anomalies in
Type~II.  A D1-brane is a compact spin submanifold $i\:W^2\hookrightarrow
X^{10}$.  It is endowed with a real vector bundle~$Q$ with connection, which
as usual we write as a cocycle~$\qh\in \Zh_{KO}^0(W)$.  The corresponding
contribution to the magnetic current is
  $$ \jh = u^{-4}\,i_*\qh.  $$
As usual magnetic current modifies the geometric nature of the gauge field,
so when added to the background magnetic current the appropriate modification
of~\thetag{3.45} is: 
  $$ \text{$\Bh$~is a nonflat isomorphism $\mh\longrightarrow
     \Eh + \jh$}.  $$
The electric coupling~$\pfaff \bigl(-q_{X/T}(\Bh)\bigr)$ is now the inverse
pfaffian of a nonflat isomorphism
  $$ q_{X/T}(\mh) \;\;\longrightarrow \;\;q_{X/T}(\Eh + \jh) \;\cong\;
     q_{X/T}(\Eh) + q_{X/T}(\jh) + B_{X/T}(\Eh,\jh).  $$
Recall that $q_{X/T}$~is defined in~\thetag{3.40} and
$B_{X/T}(\ah,\ah')=\int_{X/T} u^6\,\ah\cdot \ah'$ for $\ah,\ah'\in
\Zh^0_{KO}(X)$.  Thus the new contribution to the anomaly
(beyond~\thetag{3.47}) is 
  $$ \pfaff\,-\bigl\{q_{X/T}(\jh) + B_{X/T}(\Eh,\jh)\bigl\}. \tag{3.53} $$

Next, we rewrite the expression in braces as an integral over~$W$.  The second
term is immediate using the push-pull formula: 
  $$ B_{X/T}(\Eh,\jh) = u^2\int_{W}i^*E\cdot \qh. \tag{3.54} $$
For the first term we claim, and prove in \theprotag{B.40} {Proposition},
that
  $$ q_{X/T}(\jh) = u^2\int_{W} \Delta ^+(\nu )\cdot \lambda ^2(\qh) \;-\;
     \Delta ^-(\nu )\cdot \Sym^2(\qh), \tag{3.55} $$
where $\nu \to W$ is the normal bundle and $\Delta ^{\pm}$~are the half-spin
bundles.  

The low energy theory on the D1-brane~$W$ has fermions\footnote{I warmly
thank Jacques Distler for computing the low energy theory on Type~I
D-branes.} whose anomaly exactly cancels~\thetag{3.53}.  Namely, the D1-D9
strings give massless positive chirality spinor fields with coefficients in
the bundle $\Hom(Q,i^*E)$, and the D1-D1 strings give positive chirality
spinor fields with coefficients in $\Delta ^+(\nu )\otimes {\tsize\bigwedge}
^2(Q)$ as well as negative chirality spinor fields with coefficients in
$\Delta ^-(\nu )\otimes \Sym^2(Q)$.  As before, a geometric form of the
Atiyah-Singer index theorem computes the anomaly to be the pfaffian
of~\thetag{3.54} times the pfaffian of~\thetag{3.55}, and this cancels the
anomaly~\thetag{3.53} from the electric coupling.
 
The story for the D5-brane is parallel, except that $Q$~is quaternionic.

 \newpage
 \head
 \S{A} Appendix:  Wick Rotation
 \endhead
 \comment
 lasteqno A@  5
 \endcomment

We include these well-known remarks since some of these issues caused the
author confusion from which we hope to spare others.  The exposition has no
pretense to rigor.  
 
On Minkowski spacetime there are both classical and quantum versions of field
theory.  A classical theory consists of a symplectic manifold of fields
(often there are classical field equations which define it) and a symplectic
action of the Poincar\'e group.  A quantum theory consists of a Hilbert space
and operators (in particular for the Lie algebra of the Poincar\'e group),
and from this data one defines correlation functions of operators.  Such
classical and quantum theories exist as well on Lorentzian spacetimes of the
form~$\RtN$, where $(N,g_N)$~is Riemannian and $\RR_t$~represents time; the
Lorentz metric is~$dt^2-g_N$.  {\it Wick rotation\/} occurs in the quantum
theory as follows.  Correlation functions depend on the positions $(t,n)\in
\RtN$ of local operators.  Assuming\footnote{In axiomatic formulations of
quantum field theory on Minkowski spacetime (see~\cite{K,\S2}, for example)
this is a consequence of more basic axioms.}  they extend to holomorphic
functions of~$t$, one restricts to purely imaginary values
of~$t=\sqrt{-1}\,\tau $ to obtain correlation functions on~$\RetN$, which is
a Riemannian manifold with metric~$d\tau ^2 + g_N$.  In good cases one can
functorially define correlation functions on all Riemannian manifolds~$X$ (of
fixed dimension) not necessarily of the form~$\RetN$.  This, then, is the
substance of Euclidean field theory: quantum correlation functions.  There is
not a Hilbert space interpretation, nor does one try to make physical sense
of classical field theory.
 
In the Euclidean context one often introduces ``classical fields'' and an
action functional and then writes correlation functions as a functional
integral over fields.  We review that process briefly, but issue the warning
that despite the language of classical fields one is {\it not\/} doing
classical field theory.\footnote{In Euclidean field theory the Euler-Lagrange
equations do not have a classical meaning.  For example, the Euler-Lagrange
equation derived from the action~\thetag{2.17} is
  $$ d*F_A = -\sqrt{-1}\,C_X\,j_E,  $$
for some real form $C_X$.  As both $F_A$~and $j_E$~are real (see below), this
equation clearly has no solutions with nonzero~$j_E$.  On the other hand,
solutions to the Euclidean Euler-Lagrange equations---called {\it
instantons\/}---are relevant to the asymptotic analysis of the Euclidean
functional integral; our point here is that there is no strictly {\it
classical\/} interpretation.  We should also point out that variational
equations do have a distinguished place in Riemannian geometry.}
 
The Euclidean functional integral on ~$\RetN$ is derived formally from a
corresponding (formal) functional integral in the Lorentzian theory on the
spacetime~$\RtN$ in case the quantum theory on~$\RtN$ is obtained by
quantizing a classical theory which has a lagrangian description.  The
derivation of the Lorentzian functional integral is given in standard
texts.\footnote{One subtlety, which already occurs in quantum mechanics
($N=pt$), is that the action should be at most quadratic in the first time
derivative of the fields (see~\cite{R,p\.163}, for example).}  The main
result has the schematic form
  $$ \langle \Cal{O}  \rangle = \iint_{}\,D\phi \,D\psi \;e^{iS(\phi ,\psi
     )}\,\Cal{O}(\phi ,\psi ), \tag{A.1} $$
where 
  $$ S(\phi ,\psi ) = \int_{\RtN}L(\phi ,\psi ). \tag{A.2} $$
In these formulas $\phi $~stands for a collection of Bose fields and $\psi
$~for a collection of Fermi fields.  The fields are defined on~$\RtN$, and
in~\thetag{A.1} the integral is over fields with finite action.  It is an
important principle of unitary quantum field theory that the fields and
action are {\it real\/}.  (Complex or quaternionic notation may be used, but
the point remains.)  In the quantum theory this leads to the fact that
operators corresponding to real observables are symmetric.  The integrand
which defines the action~$S$ is the lagrangian density~$L$.  The
symbol~$\Cal{O}$ in~\thetag{A.1} denotes a functional (or finite product of
functionals) of the fields.  The left-hand side~$\langle \Cal{O} \rangle$ is
the quantum correlation function.  Now a typical operator
  $$ \Cal{O}_{(t,n)}\phi =\phi (t,n) \tag{A.3} $$
evaluates the field at a point.  It is the $t$-dependence of the left-hand
side of~\thetag{A.1} which one (formally) analytically continues to complex
values of~$t$.  We now discuss the corresponding Wick rotation of the
right-hand side.  For this one simultaneously rotates both the finite
dimensional integral in~\thetag{A.2} and the functional integral
in~\thetag{A.1}.  Bose fields and Fermi fields in~\thetag{A.1} are treated
differently.  We emphasize that our presentation is formal and algebraic.  In
any case the Lorentzian functional integral~\thetag{A.1} is usually
oscillatory and badly behaved, and one should view this heuristic argument as
motivation for the definition of correlation functions by Euclidean
functional integrals.\footnote{As Coleman~\cite{C,p\.148} says, the
Lorentzian functional integral is ``ill-defined, even by our sloppy
standards.''}
 
First, we discuss Wick rotation in the space of fields.  Partially complexify
the fields to spaces of complex-valued functions which depend holomorphically
on a complex variable~$t$.  Real fields on~$\RtN$ satisfy the reality
condition
  $$ \phi (\bar{t},n) = \overline{\phi (t,n)}. \tag{A.4} $$
Operators such as~\thetag{A.3} extend to complex operators defined on
complexified fields; they also depend holomorphically on~$t$.  The
restriction of complexified fields to purely imaginary values
of~$t=\sqrt{-1}\,\tau $, i.e., to~$\RetN$, we term {\it Euclidean fields\/}.
Here is where the treatment of Bose and Fermi fields differs.  There is no
reality constraint imposed on Euclidean Fermi fields~$\psi _E$: the fermionic
functional integral is algebraic, and extending the coefficients from~$\RR$
to~$\CC$ does not affect the answer.  For example, the pfaffian of a real
operator equals the pfaffian of its complexification.  For Euclidean Bose
fields~$\phi _E$, on the other hand, \thetag{A.4}~is ``rotated'' to the
reality constraint
  $$ \phi _E(-\bar{t},n) = \overline{\phi _E (t,n)}. \tag{A.5} $$
In particular, Euclidean Bose fields are real-valued on~$\RetN$.  This
rotation in the complexified function space is the change of integration
domain from the Lorentzian functional integral to the Euclidean functional
integral. 
 
In the finite dimensional integral~\thetag{A.2} rotate the domain of
integration from~$\RtN$ to~$\RetN$.  The Euclidean lagrangian density~$L_E$
on~$\RetN$ is defined by analytic continuation from~$L$ as a function of
Euclidean fields:
  $$ L_E(\phi ,\psi ) = \frac{1}{\sqrt{-1}} L(\phi ,\psi ).  $$
The reality constraint~\thetag{A.5} is not used in defining~$L_E$; one
essentially substitutes~$t=\sqrt{-1}\,\tau $ into~$L$ and divides
by~$\sqrt{-1}$.  This has the usual effect of changing the sign of potential
energy terms, etc.\footnote{See~\cite{DF,\S7} for typical examples.  However,
be warned that there is a crucial notational mistake in the second paragraph:
the analytic continuation of a real scalar field on Minkowski spacetime is
not real when restricted to Euclidean space.  This confusion---which extends
to more complicated fields---was one of our motivations to include this
appendix.}  Notice that imposing~\thetag{A.5} does {\it not\/} render~$L_E$
real in general.  The integral of~$L_E$ over~$\RetN$ is the Euclidean
action~$S_E$.  By a similar change of variables, operators~$\Cal{O}$ are
re-expressed as Euclidean operators~$\Cal{O}_E$.
 
With this understood the Euclidean correlation functions are
  $$ \langle \Cal{O}_E \rangle_E = \iint_{}\,D\phi_E \,D\psi_E
     \;e^{-S_E(\phi_E ,\psi_E )}\,\Cal{O}_E(\phi_E ,\psi_E ).  $$
The functional integral is over Euclidean fields of finite Euclidean action.
 
Usually the Euclidean theory may be formulated on Riemannian manifolds~$X$
not necessarily of the form~$\RetN$.  The reality condition on the classical
fields and action in the Lorentzian case is replaced by (i)\ the reality
condition on Euclidean Bose fields, and (ii)\ the condition that the
Euclidean action undergo complex conjugation when the orientation is
reversed.\footnote{Orientation reversal should be construed locally so that
the condition makes sense on unorientable manifolds.}  The corresponding
reality condition on the quantum Euclidean correlation functions is called
{\it reflection positivity\/}.  (See~\cite{DEFJKMMW,p\.690} for a formal
derivation of reflection positivity from the reality condition~(ii) on the
Euclidean action.)

 \newpage
 \head
 \S{B} Appendix: $KO$ and Anomalies in Type~I \\ \\
Daniel S. Freed\\Michael J. Hopkins
 \endhead
 \comment
 lasteqno B@ 48
 \endcomment

In this appendix we provide proofs of several assertions needed in the
anomaly cancellation arguments for Type~I given in~\S{3}.  We begin in
\theprotag{B.1} {Proposition} with a special computation of a pfaffian line
bundle, stated before~\thetag{3.39}.  Then we turn to the theory of the
quadratic form~\thetag{B.31} in differential $KO$-theory which is relevant to
the self-dual field in Type~I.  It is not symmetric about the origin; rather,
the symmetry involves a (differential) $KO$-theoretic analog of the Wu class.
We restrict to situations where this center exists.  In \theprotag{B.40}
{Proposition} we carry out a computation needed for D-brane anomaly
cancellation in Type~I.  At the end of the appendix we prove that the Adams
operation~$\psi _2$ deloops once, a fact used earlier in this appendix.  We
do not resolve all questions about the quadratic form---for example,
existence and uniqueness questions concerning the center---so view this
account as provisional.

        \proclaim{\protag{B.1} {Proposition}}
 Let $X\to T$ be a Riemannian spin fiber bundle with fibers smooth closed
spin 10-manifolds.  Let $\scrL$~be the pfaffian line bundle of the family of
Dirac operators on the fibers of~$X$ coupled to
  $$ {\tsize\bigwedge} ^2\bigl(T(X/T)+22 \bigr) + T(X/T) -2. \tag{B.2} $$
Then $\scrL$, together with its natural metric and connection, is trivial. 
        \endproclaim

\noindent 
 The natural connection is defined, and the curvature and holonomy computed,
in~\cite{BF}.  We do not claim here to give a {\it canonical\/}
trivialization, so as explained at the end of the introduction while this
proposition can be used to prove the cancellation of anomalies, it is not
strong enough to construct correlation functions.

        \demo{Proof}
 The proof is similar to the proof of \cite{FW,Theorem~4.7}. 
 
The curvature of~$\scrL$ was computed in~\thetag{3.37}, and was seen to
vanish.  To compute the holonomy we pull back over a loop in~$T$, so consider
a family $X\to S^1$.  Let the circle have its bounding spin structure; then
$X$~is a closed spin 11-manifold.  The holonomy of~$\scrL$ around~$\cir$
is\footnote{The extra factor of~2 is due to the pfaffian (as opposed to a
determinant); the absence of an adiabatic limit is due to the vanishing of
the curvature.} $\exp(-\tp i\xi _X/2)$, where $\xi _X=\frac 12(\eta _X +
\dim\ker D_X)$ is the Atiyah-Patodi-Singer invariant for the Dirac operator
coupled to~\thetag{B.2}.  Now any closed spin 11-manifold is the boundary
$X=\partial M$ of a compact spin 12-manifold, and we take~$M$ to have a
Riemannian metric which is a product near the boundary.  We compute~$\xi
_X/2\pmod1$ via the Atiyah-Patodi-Singer theorem, but for that we need to
extend~\thetag{B.2} to~$M$, which is not necessarily fibered over~$\cir$.  A
short computation shows that 
  $$ \Cal{E} = {\tsize\bigwedge} ^2(TM+20) + TM -4  $$
restricts on~$\partial M$ to~\thetag{B.2}.  Then 
  $$ \xi _X/2 \equiv \int_{M}\[\frac 12
     \Ahat(M)\,\ch(\Cal{E})\]_{(12)}\pmod1, $$
and a straightforward computation shows that the integrand vanishes.  Hence
the holonomy of~$\scrL$ is trivial.
        \enddemo

As a preliminary to the rest of this appendix we recall some facts about
$KO$-theory.  Let $X$~be any manifold.  First, there is a canonical
filtration, the {\it Atiyah-Hirzebruch filtration\/}.  A class $a\in KO(X)$
has filtration~$q$ if $q$~is the largest integer such that the pullback
of~$a$ via any smooth map $\Sigma ^{k}\to X$ of a $k$-dimensional
manifold~$\Sigma $ into~$X$ vanishes for all~$k<q$.  The product of classes
of filtration~$q$ and filtration~$q'$ has filtration~$\ge q+q'$.  Second,
suppose $\nu \to X$ is a real spin bundle of even rank~$2r$.  The {\it
$K$-theory Thom class\/}~$U$ is an element in~$K^{2r}_{\text{cv}}(\nu )$,
where `cv'~denotes `compact vertical support.'  Let $i\:X\hookrightarrow \nu
$ be the zero section.  By the splitting principle we formally write
  $$ \nu \otimes \CC = \bigoplus\limits_{i=1}^r\; (\ell _i + \ell _i\inv
     ). \tag{B.3} $$
Then 
  $$ i^*U = u^r\,\[ \Delta ^+(\nu ) - \Delta ^-(\nu )\] =
     u^r\,\prod\limits_{i=1}^r\; (\ell _i^{1/2} - \ell _i^{-1/2}), \tag{B.4}
     $$
where $\Delta ^{\pm}$~are the half-spin representations.  If $r\equiv
0\pmod4$, then $U$~is real; if $r\equiv 2\pmod4$, then $r$~is quaternionic.
When $r\equiv 0\pmod4$ there is a $KO$-theory Thom class (also called~`$U$')
whose complexification is the $K$-theory Thom class.  Next, we state the form
of Poincar\'e duality in $KO$-theory which we need.  Let $X$~be a spin
manifold of dimension $n$.  Then
  $$ \Hom\bigl(KO^q_c(X;\RZ),\RR/\ZZ \bigr) \cong KO^{n+4-q}(X), \tag{B.5} $$
where `c'~denotes `compact support.'  The analogous statement in ordinary
cohomology does not have the shift by~4.  Ordinarily, a generalized
cohomology theory does not contain such a duality statement.  Finally, we
make use of the Adams operation~$\psi _2$.  It is a natural {\it ring
endomorphism\/} of~$KO^0(X)$ for any manifold~$X$, and is related to the
exterior square~$\lambda ^2$ (which is not a ring homomorphism) by
  $$ 2\lambda ^2(a) = a^2 - \psi _2(a). \tag{B.6} $$
The operation~$\psi _2$ is defined on line bundles~$\ell $ by the
formula~$\psi _2(\ell )=\ell ^2$; it extends to arbitrary elements of~$KO$
using the splitting principle and the fact that $\psi _2$~is a ring
homomorphism.  The same definitions for $\psi _2$ work on complex $K$-theory,
and then $\psi _2$~extends to~$K^{-q}$ for~$q\ge 0$.  Its action on the Bott
element~$u\inv $ is\footnote{Since $u\inv =H-1$ for $H$~the hyperplane bundle
on~$S^2\cong \CP^1$, we compute $\psi _2(u\inv )=H^2-1 = 2(H-1)$ in the
reduced $K$-theory of~$S^2$, which is isomorphic to~$K^{-2}(pt)$.}
  $$ \psi _2(u\inv ) = 2u\inv . \tag{B.7} $$
If we invert~2, then $\psi _2$~extends to~$K^q$ for all~$q$.  Furthermore,
after inverting~2 there is an inverse operation~$\psh$.  If $\ell $~is a line
bundle, and $x=1-\ell $, then 
  $$ \psh(\ell ) = (1-x)^{1/2} = 1 - \frac 12 x - \frac 18 x^2 + \dots
     .  $$
Note that $x^q$ has filtration~$\ge q$, so on a finite dimensional manifold
the infinite series terminates.  On an infinite dimensional manifold we must
work in a certain completion~$\Khat$ of $K$-theory.  We also need a single
delooping of~$\psi _2$. 

        \proclaim{\protag{B.8} {Proposition}}
 There exists an operation $\psi _2\:KO^1(X)\to KO^1(X)$ which is compatible
with the standard~$\psi _2$ under suspension, i.e., the diagram 
  $$ \CD 
      KO^0(X) @>\psi _2>> KO^0(X)\\ 
      @VVV @VVV\\ 
      KO^1(\Sigma X) @>\psi _2>> KO^1(\Sigma X)\endCD \tag{B.9} $$
commutes.  Furthermore, $\psi _2$~extends to an operation on the differential
group~$(\KOh)^1(X)$ and so restricts to an operation on~$KO^0(X;\RZ)$. 
        \endproclaim

\flushpar
 The proof, which involves homotopy-theoretic techniques, is deferred to the
end of the appendix.  More elementary is the extension of the
operations~$\psi _2$ and~$\lambda ^2$ to the differential group
$(\KOh)^{0}$.  For that we use the fact that the topological operations are
defined on the level of cochains, not just cohomology, and there are
compatible operations on differential forms.  Then such operations on~$\KOh$
are defined from the basic pullback square~\thetag{1.5}.

Concerning the Atiyah-Hirzebruch filtration, we have the following easy
statement.

        \proclaim{\protag{B.10} {Lemma}}
 Suppose $F\to X$ is a real vector bundle with $w_1(F)= w_2(F)=0$.  Then the
class of~$(F - \rank F)$ in~$KO(X)$ has filtration~$\ge 4$. 
        \endproclaim

        \demo{Proof}
  The classifying map $X\to \ZZ \times BO$ of~$(F - \rank F)$ lifts
to~$B\Spin$, and the 3-skeleton of~$B\Spin$ is trivial.
        \enddemo

We introduce a characteristic class~$\ph(F)\in \widehat{KO}[\frac 12](X)$
associated to a real vector bundle $F\to X$.  Let $i\:X\to F$ be the zero
section.  Suppose first that $F$~is spin of even rank~$2r$.  Let~$\ell
_i,\,i=1,\dots ,r$ as in~\thetag{B.3} and set
  $$ \ph(F) = i^*\(\frac{U}{\psh(U)}\) = \prod\limits_{i=1}^r\,\frac{\ell
     _i^{1/4} + \ell _i^{-1/4}}{2}, \tag{B.11} $$
where we use~\thetag{B.4} and~\thetag{B.7}.  Here $\psh(U)\in
\widehat{KO}[\frac 12]^{2r}_{\text{cv}}(F)$ and under the Thom isomorphism it
corresponds to a class $i^*\bigl(\psh(U) \bigm/U\bigr)\in \widehat{KO}[\frac
12]^{0}(X)$ which has the form $1+z$ for $z$~of filtration~$\ge 1$.  The
characteristic class~$\ph(F)$ is its inverse.  From the last expression
in~\thetag{B.11} we see that $\ph(F)$~is defined for any real vector
bundle~$F$.  To compute a formula for~$\ph(F)$, write~$\ell _i = 1-x _i$ and
$\ell _i\inv =1-y_i$.  Expand~\thetag{B.11} using the binomial theorem, take
the log, and write the result in terms of~$s_i = x_i + y_i = x_iy_i$ using
the Newton polynomials for~$x_i^n+y_i^n$:
  $$ \log \ph(F) = \sum\limits_{i=1}^r @,@,@,(-\frac{1}{32}@,@,@,s_i -
     \frac{3}{1024}@,@,@,s_i^2 - \frac{5}{12288}@,@,@,s_i^3 + \cdots).  $$
Note $\sum s_i$~is the reduced bundle~$2r-F$, so has filtration~$\ge 1$.  We
then express power sums in~$s_i$ in terms of the elementary symmetric
polynomials~$p _1,p _2,\cdots $ in~$s_i$ and exponentiate:
  $$ \ph(F) = 1 - \left( \frac{1}{32}@,@,@,p _1 \right) + \left (
     \frac{3}{512}@,@,@,p _2 - \frac{5}{2048} @,@,@,p _1^2 \right) + \left(
     -\frac{5}{4096}@,@,@,p_3 + \frac{17}{16384}@,@,@,p_1p_2 -
     \frac{21}{65536}@,@,@,p_1^3 \right)+ \cdots .  \tag{B.12} $$
Finally, we compute 
  $$ \aligned
      p _1 &= 2r - F \\
      p _2 &= (2r^2 - 3r) - (2r-2)@,@,@,F + \lambda ^2F \\
      p _3 &= \left( \frac{4r^3 - 18r^2 + 20r}{3} \right) -
     (2r^2+7r-5)@,@,@,F + 2(r-2)@,@,@,\lambda ^2F - \lambda ^3F,\endaligned $$
so find 
  $$ \align
      \ph(F) = 1 &+ \frac{1}{2^5}[F - 2r] \\
      &\;\;\;+ \frac{1}{2^{11}}\Bigl[7\lambda ^2(F) - 5\Sym^2(F) + (24-4r)F +
     (4r^2 - 36r) \Bigr]\\
      \vspace{3pt}
      &\;\;\;\;\;\;+\frac{1}{2^{16}}\Biggl[ 33\lambda ^3(F) -26R(F)+ 21
     \Sym^3(F) -(14r-184)\lambda ^2F + (10r - 136)\Sym^2(F) \\
      \vspace{-12pt}
      &\qquad\qquad \qquad + (4r^2 + 1036r - 400)F - \left ( \frac{8r^3 -
     216r^2 + 1600r}{3} \right)\Biggr] + \cdots . \tag{B.13} \endalign $$
Here $R(F)$~is the associated bundle to~$F$ which satisfies 
  $$ F^{\otimes 3} \cong \Sym^3F \oplus 2R(F) \oplus \lambda ^3F,\qquad
     F\otimes \lambda ^2F \cong  \lambda ^3F \oplus R(F).  $$
Note from \theprotag{B.10} {Lemma} and \thetag{B.12} that if
$w_1(F)=w_2(F)=0$, then the second term in~\thetag{B.13} has filtration~$\ge
4$, the third term has filtration~$\ge 8$, etc.

For a finite dimensional manifold~$Y$ define $\ph(Y)=\ph(TY)\in
KO(Y)[\frac{1}{2}]$.  

        \proclaim{\protag{B.14} {Proposition}}
 Let $Y^{8n+4}$ be a spin manifold.  Then $2^{4n+2}\ph(Y)$ is the image
in~$KO[\frac 12](Y)$ of a canonical class~$\lambda _n(Y)\in KO(Y)$. 
        \endproclaim

\flushpar 
  The class~$\lambda _n(Y)$ is a $KO$-theoretic analog of a Wu class.  It is
defined for any spin manifold~$Y^d$ of dimension~$\le 8n+4$; apply
\theprotag{B.14} {Proposition} to~$Y^d\times \RR^{8n+4-d}$.

        \demo{Proof}
 By Poincar\'e duality~\thetag{B.5} the functional 
  $$ \aligned
      KO^0_c(Y;\RZ) &\longrightarrow KO^{-(8n+4)}(pt;\RZ)\cong \RZ \\ 
      a &\longmapsto \int_{Y}\psi _2(a)\endaligned   \tag{B.15} $$
is represented by a class $u^{4n+4}\lambda _n(Y)\in KO^{8n+8}(Y)$: 
  $$ \int_{Y}\psi _2(a) = \int_{Y}\lambda _n(Y)\cdot a,\qquad a\in
     KO_c^0(Y;\RZ). \tag{B.16} $$
Note that the existence of the functional~\thetag{B.15} relies on
\theprotag{B.8} {Proposition}.  Now integration in $KO$-theory is defined
using an embedding $i\:Y\hookrightarrow \RR^N$ with $N=(8n+4) + 8k$ for
some~$k$; then
  $$ i_*\:KO^0(Y;\RZ)\longrightarrow KO^{8k}_c(\RR^N;\RZ)\cong \RZ 
     $$
is the integral.  Let $U$~be the $KO$~Thom class of the normal bundle $\nu
\to Y$ of~$Y$ in~$\RR^N$.  Then we compute 
  $$ \split
      i_*\psi _2(a) &= \psi _2(a)\cdot U \\ 
      &= \psi _2\bigl(a\cdot\psh(U) \bigr) \\ 
      &= 2^{-4k}a\cdot\psh(U) \\ 
      &= 2^{-4k}\( a\cdot\frac{\psh(U)}{U}\)\cdot U.\endsplit \tag{B.17} $$
In the first equation we pull~$\psi _2(a)$ back to~$\nu $ and extend $\psi
_2(a)\,U$ to~$\RR^N$ using the fact that $U$~has compact vertical support.
In the second equation we regard~$U$ in $KO$~with 2~inverted.  In the third
equation we use the fact that $KO^{8k}_c(\RR^N)$~is generated by~$u^{-4k}$.
Thus from~\thetag{B.7} we know that $\psi _2$~acts on~$KO_c^{8k}(\RR^N)$
and~$KO_c^{8k}(\RR^N;\RR)$ as multiplication by~$2^{-4k}$; it now follows
from ~\thetag{B.9} that $\psi _2$~also acts on~$KO_c^{8k}(\RR^N;\RZ)$ as
multiplication by~$2^{-4k}$.  Let $V$~be the $KO$~Thom class of $TY\to Y$.
Since $TY\oplus \nu $ is trivial of rank~$N$, we deduce
  $$ \frac{\psh(U)}{U}\,\frac{\psh(V)}{V} = \frac{\psh(u^N)}{u^N} =
     2^{N/2}.  $$
This is an equation in~$KO(Y)$; the factors on the left-hand side are
implicitly restricted to ~$Y$.  Substituting into~\thetag{B.17} we find 
  $$ i_*\psi _2(a) = a\cdot \(2^{4n+2}\,\frac{V}{\psh(V)}\)\cdot U =
     i_*\bigl(a\cdot 2^{4n+2}\ph(V) \bigr). \tag{B.18} $$
Thus 
  $$ \int_{Y}\psi _2(a) = \int_{Y}2^{4n+2}\ph(Y)\cdot a,\qquad a\in
     KO_c^0(Y;\RZ). $$
Comparing with~\thetag{B.16}, and using the Poincar\'e duality
isomorphism~\thetag{B.5}, we deduce that the image of~$\lambda _n(Y)$
in~$KO[\frac 12](Y)$ is~$2^{4n+2}\ph(Y)$, as desired. 
        \enddemo

On a spin manifold of dimension~$\le 8n+3$, the class~$\lambda _n$ is
canonically divisible by~2.  (Compare with a similar assertion about Wu
classes in~\cite{HS}.)

        \proclaim{\protag{B.19} {Proposition}}
 Let $X^{8n+3}$ be a spin manifold.  Then there is a canonically associated
class $\mu _n(X)\in KO(X)$ with $2\mu _n(X)=\lambda _n(X)$.  
        \endproclaim

\flushpar
 The proposition applies to manifolds of dimension~$< 8n+3$ by taking the
product with a vector space as before.  

        \demo{Proof}
 The operation~$\lambda ^2$ loops to an operator~$\Omega \lambda ^2$
on~$KO\inv (X)$.  It is linear since products of suspended classes vanish.
Similarly, there is a linear operator~$\Omega \lambda ^2$ on~$KO^{-1}(X;\RZ)$
compatible with~$\Omega \lambda ^2$ on~$KO\inv (X;\RR)$ and~$\lambda ^2$
on~$KO^0(X)$ in the long exact sequence.  From~\thetag{B.6} we have
  $$ 2\,\Omega \lambda ^2 = -\psi _2. \tag{B.20} $$
Now Poincar\'e duality~\thetag{B.5} implies that the linear functional
  $$ \aligned
      KO^{-1}_c(X;\RZ) &\longrightarrow KO^{-(8n+4)}(pt;\RZ)\cong \RZ \\
      a &\longmapsto \int_{X}\Omega \lambda ^2(a )\endaligned
      $$
is represented by a class $-u^{4n+4}\mu _n(X)\in KO^{8n+8}(X)$: 
  $$ \int_{X}\Omega \lambda ^2(a ) = \int_{X}-\mu _n(X)\cdot a,\qquad a\in
     KO_c^{-1}(X;\RZ).  \tag{B.21} $$
From~\thetag{B.20} and~\thetag{B.16} we have 
  $$ \int_{X}2\,\Omega \lambda ^2(a ) = \int_{X}-\psi _2(a ) = 
     \int_{X}-\lambda _n(X)\cdot a. \tag{B.22} $$
Comparing~\thetag{B.22} and~\thetag{B.21} we deduce $2\mu _n(X)=\lambda
_n(X)$.  
        \enddemo

Turning to {\it differential\/} $KO$-theory we have the following. 

        \proclaim{\protag{B.23} {Proposition}}
 Let $Y^{8n+4}$ be a Riemannian spin manifold.  Then there is a canonical
lift $\lh_n(Y)\in (\KOh)^0(Y)$ of~$\lambda _n(Y)$ such that 
  $$ \int_{Y}\psi _2(\ah) = \int_{Y}\lh_n(Y)\cdot \ah\quad \in \RZ 
     $$
for all $\ah\in (\KOh)^1_c(Y)$. 
        \endproclaim

\flushpar
 The proof is parallel to the proof of \theprotag{B.14} {Proposition}.  It
relies on Poincar\'e duality for~$\KOh$, which on an $n$-dimensional
Riemannian spin manifold~$Y$ states that there is an ``almost perfect''
pairing 
  $$ \alignedat2
      (\KOh)^{q+1}_c(Y)&\otimes (\KOh)^{n+4-q}(Y) &&\longrightarrow \RZ \\
      \ah \quad \quad &\otimes\quad \quad \bh &&\longmapsto \int_{Y}\ah\cdot
     \bh.\endaligned $$
Note that the integral lands in $(\KOh)^5(pt)\cong KO^{4}(pt;\RZ)\cong \RZ$.
This duality combines the topological duality~\thetag{B.5} with a duality on
differential forms---see~\thetag{1.8} and~\thetag{B.18}.  The ``almost
perfect'' refers to the fact that the dual of a differential form is a de
Rham current.  Thus the application of $\KOh$-theory Poincar\'e duality to
the functional $\ah\mapsto \int_{Y}\psi _2(\ah),\;\ah\in (\KOh)^1_c(Y)$ only
gives a distributional class~$\lh_n(Y)$.  Then the computation of its image
in $\bigl(KO[\frac 12]\spcheck \bigr)(Y)$, parallel to the computation in the
proof of \theprotag{B.14} {Proposition}, shows that its curvature is in fact
smooth.  
 
There is no {\it canonical\/} lift of~$\mu _n(X)$ to a differential class on a
Riemannian spin $(8n+3)$-manifold, though lifts do exist.  We define suitable
lifts below.

Specialize to~$n=1$, so to the classes $\lambda (Y) = \lambda _1(Y)\in KO^0(Y)$
and $\lh(Y)=\lh_1(Y)\in (\KOh)^0(Y)$ canonically associated to a Riemannian
spin 12-manifold~$Y$.  We also have the topological class $\mu (X)=\mu
_1(X)\in KO^0(X)$ canonically associated to a spin 11-~or 10-manifold~$X$.
Note by~\thetag{B.13} that
  $$ \lambda (Y) = 2(TY+22) + (\text{classes of filtration $\ge 8$}) \qquad
     \in KO[\tfrac 12]^0(Y);  $$
there are similar equations for~$\lh(Y)$ and~$\mu (X)$ after inverting~2.   

        \proclaim{\protag{B.24} {Lemma}}
 Let $Y$~be a Riemannian spin 12-manifold and $X$~a Riemannian spin 11-~or
10-manifold.  Then \rom(without inverting~2\rom) 
  $$ \align
      \lambda (Y) &= 2(TY + 22) + \epsilon _1 \tag{B.25}\\
      \lh(Y) &= 2(\Th Y+22) + \eh_1  \tag{B.26} \\ 
      \mu (X) &= \phantom{2}(TX + 22) + \epsilon _2 \tag{B.27} \endalign $$ 
where $\epsilon _1,\eh_1,\epsilon _2$~have filtration $\ge 8$. 
        \endproclaim

\flushpar
 Here $\Th Y\in (\KOh)^0(Y)$ is the class of the tangent bundle of~$Y$ with
its Levi-Civita connection.  Also, we induce a filtration on $\KOh(X)$ from
the Atiyah-Hirzebruch filtration on~$KO(X)$ via the characteristic class
$\KOh(X)\to KO(X)$.   

        \demo{Proof}
The first assertion~\thetag{B.25} is equivalent to
  $$ f(Y,a) := \int_{Y}\psi _2(a) - 2(TY+22)a = 0,\qquad a\in
     KO^0(Y;\RZ),\quad \filt a\ge8. \tag{B.28} $$
(Note that an element of~$KO^0(Y;\RZ)$ of filtration $\ge5$ has filtration
$\ge8$.)  There is a similar rewriting of the other two assertions.  As a
first step we argue that it suffices to assume that $Y$~is compact.  Namely,
using a proper Morse function we can find a compact manifold with
boundary~$Y'$ contained in~$Y$ such that the support of~$a$ lies in the
interior of~$Y'$; then replace $Y$~by the double of~$Y'$.  This does not
change the value of~\thetag{B.28}.  Then our proof of~\thetag{B.28} relies on
the fact that $f(Y,a)$~depends only on the bordism class of~$(Y,a)$: If
$Y^{12} = \partial Z^{13}$ for a compact spin manifold~$Z$, and $a$~extends
to a class on~$Z$, then \thetag{B.28}~vanishes.  Since $f(Y,0)=0$ it suffices
to consider the bordism class of~$(Y,a) - (Y,0)$.  The spectrum which
classifies such reduced pairs is~$\MSB$; an element of~$\pi _n(\MSB)$
represents a spin $n$-manifold which bounds together with a class in~$KO^0$
of filtration~$\ge 8$.  One computes
  $$ \aligned
      \pi _{10}(\MSB)&\cong \zt, \\ 
      \pi _{11}(\MSB)&=0, \\ 
      \pi _{12}(\MSB)&\cong \ZZ\times \ZZ, \\ 
      \pi _{13}(\MSB)&=0.\endaligned \tag{B.29} $$
From these facts one deduces that the reduced bordism group of pairs~$(Y,a)$
with $a\in KO^0(Y;\RZ)\langle8\rangle$ is isomorphic to~$\RZ\times \RZ$.  In
particular, it is arbitrarily divisible.  Since $f(Y,32b)=0$ for all~$b$
(see~\thetag{B.13}), and any $(Y,a)-(Y,0)$ is bordant to~$(Y,32b) - (Y,0)$
for some~$b$ by the divisibility, we obtain the desired result~$f(Y,a)=0$.
 
The proof of~\thetag{B.27} is similar; the relevant bordism group of reduced
pairs $(X,a) - (X,0),\;a\in KO^{-1}(X;\RZ)\langle8\rangle$ is again isomorphic
to~$\RZ\times \RZ$. 
 
For~\thetag{B.26} we must show 
  $$ g(Y,\ah) := \int_{Y}\psi _2(\ah) - 2(\Th Y+11)\cdot \ah  $$
vanishes for all $\ah\in (\KOh)^1(Y)$ of filtration $\ge8$.  From the exact
sequence~\thetag{1.7}, the topological result~\thetag{B.25}, and the fact
that differential forms are divisible we conclude that $g(Y,\ah)$ depends
only on the characteristic class~$[a]\in KO^1(Y)$ of~$\ah$.  But the
assertion about~$\pi _{11}$ in~\thetag{B.29} implies that $(Y,[a]) - (Y,0)$
vanishes in the appropriate reduced bordism group, whence $g$~vanishes. 
 
We remark that \theprotag{B.24} {Lemma} follows formally from the stronger
\theprotag{B.36} {Lemma} below using more bordism theory. 
        \enddemo

The following definition is analogous to the definition of a square root of
the canonical bundle of a Riemann surface. 

        \definition{\protag{B.30} {Definition}}
 Let $X$~be a Riemannian spin 11- or 10-manifold.  Then a {\it
$\mh$-structure\/} on~$X$ is a class $\mh(X)\in (\KOh)^0(X)$ such that
 \roster
 \item "\rom(i\rom)"\ $2@,@,@,\mh(X) = \lh(X)$; 
 \item "\rom(ii\rom)"\ The cohomology class of~$\mh(X)$ is~$\mu (X)$; 
 \item "\rom(iii\rom)"\ $\mh(X)$~differs from~$\Th X+22$ by an element of
filtration~$\ge 8$.
 \endroster
        \enddefinition

\flushpar
 The preceding shows that $\mh$-structures exist; differences of
$\mh$-structures are certain points of order~2 on the torus $KO\inv
(X;\RR)/ KO\inv (X)$.

Let $X\to T$ be a Riemannian spin fiber bundle with fibers closed manifolds
of dimension~10.  Recall from~\thetag{3.40} the quadratic form
  $$ \aligned
      q=q_{X/T}\: \Zh_{KO}^0(X) &\longrightarrow \Zh_{KSp}^2(T) \\
      \ah &\longmapsto \int_{X/T} u^6\,\lambda ^2(\ah)\endaligned
       \tag{B.31} $$
which refines the bilinear form 
  $$ \alignedat2
      B=B_{X/T}\: \Zh_{KO}^0(X)&\times \Zh_{KO}^0(X) &&\longrightarrow
     \Zh_{KSp}^2(T) \\
      \ah\;\quad &\times \;\quad \ah' &&\longmapsto \int_{X/T} u^6\,\ah\cdot
     \ah'.\endaligned $$
Note that $q(0)=0$.  The quadratic form~$q$ does not necessarily have a
symmetry; we restrict to fiber bundles for which a symmetry exists for the
pfaffian.  

        \definition{\protag{B.32} {Definition}}
 A {\it $\lh$-structure\/} on a Riemannian spin fiber bundle $X\to T$ of
closed 10-manifolds is a cocycle $\lh = \lh(X/T)\in \Zh_{KO}^0(X)$ and
isomorphisms 
  $$ \pfaff q(\ah) \cong \pfaff q(\lh - \ah) \tag{B.33} $$
natural in~$\ah\in \Zh^0_{KO}(X)$.
        \enddefinition

\flushpar
 An easy computation shows that \thetag{B.33}~is equivalent to natural
isomorphisms 
  $$ \pfaff \int_{X/T}\psi _2(\ah)\cong \pfaff \int_{X/T}\lh\cdot \ah
     \tag{B.34} $$
together with an isomorphism $\pfaff q(\lh)\cong 0$.  (Note
\thetag{B.34}~implies $\pfaff \,2q(\lh)\cong 0$.)  Also, \theprotag{B.23}
{Proposition} implies that the equivalence class of the restriction of~$\lh$
to the fiber is canonically determined.  A computation parallel to that in
the proof of \theprotag{B.14} {Proposition}, now for fiber bundles and in
differential~$KO$, computes the image of~$\lh$ in~$\bigl(KO[\frac 12]\spcheck
\bigr)^0(X)$ as 
  $$ \lh = 2\TTh + \frac {1}{32}\bigl( 7\lambda ^2(\Th) - 5\Sym^2(\Th) + 4\Th
     - 80 \bigr) + \cdots , \tag{B.35} $$
where $\Th=\Th(X/T)$ and~$\TTh = \Th + 22$.  More precisely, analogous to
\theprotag{B.24} {Lemma} we have the following.

        \proclaim{\protag{B.36} {Lemma}}
 $\lh\cong 2\TTh$ modulo cocycles of filtration~$\ge 8$. 
        \endproclaim

        \demo{Proof}
 As in the proof of \theprotag{B.24} {Lemma} we must show 
  $$ \pfaff \int_{X/T} \psi _2(\ah) \cong \pfaff \int_{X/T} 2(\TTh + 22)\cdot
     \ah  $$
for all~$\ah$ of filtration~$\ge 8$.  It follows from~\thetag{B.29} that the
universal family of spin 10-manifolds (up to bordism) together with a class
in~$KO$ of filtration~$\ge8$ is simply-connected.  Thus to
prove~\thetag{B.39}---an isomorphism of circle bundles with connection---it
suffices to prove that some powers are isomorphic over the universal
parameter space, since there are no flat circle bundles there
(cf.~\thetag{1.7} for~$\Hh^2$).  But this follows from~\thetag{B.35}.
        \enddemo

As for a single manifold, there is no canonical division of~$\lh$ by~2, so no
canonical center for~$q$.  We restrict to fiber bundles for which a center
exists.

        \definition{\protag{B.37} {Definition}}
 A {\it $\mh$-structure\/} on a Riemannian spin fiber bundle $X\to T$ with
$\lh$-structure is a cocycle $\mh=\mh(X/T)\in \Zh^0_{KO}(X)$ and an
isomorphism $2\mh\cong \lh$ such that $\mh=\TTh$ modulo terms of
filtration~$\ge 8$.
        \enddefinition

\flushpar
 These are the fiber bundles used in~\S{3}.  We leave to the future an
investigation of existence and uniqueness questions for $\lh$-~and
$\mh$-structures. 
 
Next, we prove a fact used in~\thetag{3.39}.

        \proclaim{\protag{B.38} {Proposition}}
 Let $X\to T$ be a fiber bundle with a $\mh$-structure, as in
\theprotag{B.37} {Definition}.  Then
  $$ \pfaff q(\mh) = \pfaff q(\TTh) =
     \pfaff\int_{X/T}{\tsize\bigwedge} ^2\TTh.  $$
        \endproclaim

        \demo{Proof}
 Set $\eh = \mh - \TTh$.  Then $\eh$~has filtration~$\ge8$, whence $\pfaff
B(\eh,\eh)=0$.  Thus
  $$ q(\TTh) = q(\mh - \eh) = q(\mh) - q(\eh) + B(\eh,\eh - \mh), 
     $$
so it suffices to prove 
  $$ \pfaff q(\eh) \cong \pfaff B(\eh,-\mh).  \tag{B.39} $$
In fact, \thetag{B.39}~holds for {\it any\/} class~$\eh$ of
filtration~$\ge8$.  As in the proof of \theprotag{B.36} {Lemma}, we must only
prove some power of~\thetag{B.39}.  Now
  $$ \pfaff B(\eh,-2\mh) \cong \pfaff B(\eh,\eh-2\mh) \cong \pfaff
     B(\eh,\eh-\lh), $$
and from ~\thetag{B.33} or~\thetag{B.34} it follows that this is isomorphic
to~$\pfaff\,2q(\eh)$, which gives the square of~\thetag{B.39}. 
        \enddemo

We now prove~\thetag{3.55}, which we restate as follows.   

        \proclaim{\protag{B.40} {Proposition}}
 Let $X\to T$ be a fiber bundle of 10-manifolds with Riemannian, spin, and
$\mh$-structures, and $W\to T$ a fiber bundle of 2-dimensional spin
submanifolds.  Denote the inclusion map as $i:W\hookrightarrow X$.  Then
for $\qh\in \Zh^0_{KO}(W)$,  
  $$ q(u^{-4}\,i_*\qh) = u^2\int_{W} \Delta ^+(\nu )\cdot \lambda ^2(\qh) \;-\;
     \Delta ^-(\nu )\cdot \Sym^2(\qh), \tag{B.41} $$
where $\nu \to W$ is the normal bundle and $\Delta ^{\pm}$~are the half-spin
bundles.   
        \endproclaim

        \demo{Proof}
 Quite generally, for any manifold~$W$ let
$\pi \:\nu \to W$ be a rank~8 real spin bundle\footnote{The computation holds
for any even rank over the complexes.  For rank~8 the half-spin bundles
associated to~$\nu $ are real; for rank~4 they are quaternionic.} and $Q\to
W$ a real vector bundle of rank~$r$.  Denote the zero section of~$\nu $ as
$i\:W\hookrightarrow \nu $.  We first compute the element~$x\in KO^0(W)$
defined by
  $$ x := u^4 \,\pi _*\lambda ^2(u^{-4}i_*Q). \tag{B.42} $$
We claim that 
  $$ x = \Delta ^+(\nu )\cdot\lambda ^2(Q) - \Delta ^-(\nu )
     \cdot\Sym^2(Q). \tag{B.43} $$
Let $U\in KO^8_{\text{cv}}(\nu )$ be the Thom class.  Then
\thetag{B.42}~implies
  $$ U\cdot\pi ^*x = u^4 \,\lambda ^2\bigl(U\cdot\pi ^*(u^{-4}Q)\bigr).  $$
Apply~$i^*$ to conclude 
  $$ i^*U\cdot x = u^4 \,\lambda ^2(i^*U\cdot u^{-4}Q). \tag{B.44} $$
This equation, and its solution~\thetag{B.43}, may be viewed as equations in
the representation ring $RSpin\mstrut _8 \times RSO\mstrut _r$; the
corresponding relations in~$KO^0(W)$ then follow by passing to the principal
bundles underlying~$\nu ,Q$ and the vector bundles associated to the
representations.  Note \thetag{B.44}~is an equation of real representations,
but we prove it by working in the complex representation ring.  To compute
the right-hand side of~\thetag{B.44} we use the Adams operation~$\psi _2$ in
the representation ring.  Use the splitting principle---i.e., restrict to the
maximal torus of~$Spin\mstrut _8$---to write $\nu \otimes \CC=
\bigoplus\limits _{i=1}^4\;(\ell \mstrut _i \oplus \ell _i\inv )$.
From~\thetag{B.4} we compute (the factors of~$u$ cancel)
  $$ \split
      \psi _2(i^*U\cdot u^{-4}Q) &= \prod\limits_{i=1}^4\;(\ell _i\mstrut
     -\ell _i\inv )\cdot \psi _2(Q) \\
      &= \prod\limits_{i=1}^4\;(\ell _i^{1/2} -\ell _i^{-1/2} )\cdot (\ell
     _i^{1/2} +\ell _i^{-1/2} )\cdot \psi _2(Q) \\
      &= \frac{i^*U}{u^4}\cdot \bigl[\Delta ^+(\nu )+\Delta ^-(\nu
     )\bigr]\cdot \psi _2(Q).\endsplit $$
Hence from~\thetag{B.6} 
  $$ \split
      2u^4\,\lambda ^2(i^*U\cdot u^{-4}Q) &= i^*U\cdot \Bigl\{ \bigl[\Delta
     ^+(\nu )-\Delta ^-(\nu )\bigr] \cdot Q^2\;-\;\bigl[\Delta ^+(\nu
     )+\Delta ^-(\nu )\bigr]\cdot \psi _2(Q)\Bigr\}\\
      &= 2\,i^*U \cdot \bigl\{ \Delta ^+(\nu )\cdot \lambda ^2(Q) \;-\;
     \Delta ^-(\nu )\cdot\Sym^2(Q)\bigr\},\endsplit \tag{B.45} $$
where we use $Q^2 = \lambda ^2(Q) + \Sym^2(Q)$.  We deduce the desired
result~\thetag{B.43} from~\thetag{B.44} and~\thetag{B.45} using the fact that
the ring $RSpin\mstrut _8\times RSO\mstrut _r$ has no zero divisors.
Finally, this universal relation in the representation ring applies to
bundles with connection, so to differential $KO$-theory, whence
\thetag{B.41}~holds. 
        \enddemo

Finally, we provide the proof of the delooping of~$\psi _2$. 

        \demo{Proof of \theprotag{B.8} {Proposition}}
 The construction is based on Atiyah's construction of Adams
operations~\cite{A}.  Start with $x\in KO^{0}(X)$ and square it, remembering
the $\ZZ/2$-action, to get $P(x)\in KO^{0}_{\ZZ/2}(X)$.  Since the group
$\ZZ/2$ is not acting on $X$, there is an isomorphism
  $$ KO^{0}_{\ZZ/2}(X) \approx RO(\ZZ/2)\otimes KO^{0}(X), $$
where 
  $$ RO(\ZZ/2)=\ZZ[t]/(t^{2}-1) $$
is the real representation ring of $\ZZ/2$.  The Adams operation is the
image of $P(x)$ in
  $$ \ZZ \underset{RO(\ZZ/2)}\to {\otimes}KO_{\ZZ/2}^{0}(X)=KO^{0}(X), $$
where the ring homomorphism $RO(\ZZ/2)\cong \ZZ[t]/(t^{2}-1)\to \ZZ$ sends~
$t$ to $-1$.

This whole discussion would make sense for $X$ a {\it spectrum},
provided we had an equivariant map $X\to X\wedge X$ to play the role
of the diagonal.  We'll define the operation $\psi_{2}$ on $KO^{1}(X)$
by defining it on $KO^{0}(S^{-1}\wedge X)$.  So start with $x\in
KO^{0}(S^{-1}\wedge X)$, form the equivariant external square, and
restrict along the diagonal $X\to X\wedge X$ to define
  $$ P(x)\in KO^{0}_{\ZZ/2}(S^{-1}\wedge S^{-1}\wedge X). $$
For $a,b\ge0$, let $S^{a+b\,t}$ be the $1$-point compactification of the
representation of $\ZZ/2$ on $\RR^{a}_{(+1)}\times \RR^{b}_{(-1)}$.  (The
subscript indicates the eigenvalue of action of the non-trivial element of
$\ZZ/2$.)  By forcing the exponents to add under smash product, we define
equivariant spectra $S^{a+b\,t}$ for all $a,b\in\ZZ$.  The shearing
isomorphism implies the sphere $S^{-1}\wedge S^{-1}$ with the flip action is
isomorphic to $S^{-1-t}$, so we can regard
  $$ P(x)\in KO^{0}_{\ZZ/2}(S^{-t}\wedge S^{-1}\wedge X). $$
We'll produce below, for any spectrum $Y$ with trivial $\ZZ/2$ action, a
canonical isomorphism 
  $$ KO^{0}_{\ZZ/2}(S^{-t}\wedge Y)\approx KO^{0}(Y). \tag{B.46} $$
In particular, this gives an isomorphism 
  $$ KO^{0}_{\ZZ/2}(S^{-1}\wedge S^{-1}\wedge X)\approx KO^{0}(S^{-1}\wedge
     X). \tag{B.47} $$
We then define
  $$ \psi_{2}(x)\in KO^{0}(S^{-1}\wedge X) $$
to be the image of $P(x)$ under~\thetag{B.47}.

To construct~\thetag{B.46} consider the cofibration
  $$ S^{0}@>>> S^{t}@>>> S^{t}\wedge \ZZ/2_{+} $$
in which the first map is map of suspension spectra gotten by
suspending the inclusion of the fixed points.   Smash this with
$S^{-t}$ to get
  $$ S^{-t}@>>> S^{0}@>>> S^{0}\wedge \ZZ/2_{+}. $$
Passing to equivariant $KO$-groups leads to a sequence
  $$ 0 @>>> \ZZ @>{1+t}>> RO(\ZZ/2) @>>>
     KO_{\ZZ/2}^{0}(S^{-t}) @>>> 0, $$
from which it follows that  
  $$ \aligned
      KO^{0}_{\ZZ/2}(S^{-t}) &\cong  RO(\ZZ/2)/(1+t) \cong  \ZZ \\
     KO^{1}_{\ZZ/2}(S^{-t}) & =0.\endaligned $$
Smashing this sequence with $Y$ then leads to a short exact sequence
  $$ 0 @>>> KO^{0}(Y) @>1+t>> RO(\ZZ/2)\otimes KO^{0}(Y) @>>>
     KO^{0}_{\ZZ/2}(S^{-t}\wedge Y) @>>> 0, $$
which gives the desired result~\thetag{B.46}.

It is useful to note that the map $S^{-t}\to S^{0}$ is also the one derived
from the diagonal map $S^{1}\to S^{1}\wedge S^{1}$ in  
  $$ \left(S^{-1}\wedge_{\text{flip}} S^{-1} \right) \wedge S^{1}
     \longrightarrow \left( S^{-1}\wedge S^{1} \right) \wedge_{\text{flip}}
     \left( S^{-1}\wedge S^{1} \right) = S^{0} \tag{B.48} $$
with the $\ZZ/2$ action as indicated.

Now suppose that $X=S^{1}\wedge Y$.  We need to show that the diagram
  $$ \CD
      KO^{0}(Y) @>\psi_{2}>> KO^{0}(Y) \\
      @VVV @VVV \\
      KO^{0}\left(S^{-1}\wedge \left(S^{1}\wedge Y \right)\right)
     @>>\psi_{2}> KO^{0}\left(S^{-1}\wedge \left(S^{1}\wedge Y \right)\right)
     \endCD $$
commutes.  The main thing to check is that the map  
  $$ \left(S^{-1}\wedge_{\text{flip}} S^{-1} \right) \wedge S^{1}
     \longrightarrow \left( S^{-1}\wedge S^{1} \right) \wedge_{\text{flip}}
     \left( S^{-1}\wedge S^{1} \right) = S^{0}, $$
derived from the diagonal map of $S^{1}$, leads to a factorization
  $$ \CD
      RO(\ZZ/2) @>>> KO^{0}_{\ZZ/2}(S^{-1}\wedge S^{-1}\wedge S^{1}) \\
      @V t \;\mapsto\; -1VV @VV\approx V \\
      \ZZ @<<< KO^{0}(S^{-1}\wedge S^{1}) \endCD $$
in which the isomorphism labeled ``$\approx$'' is the one of~\thetag{B.47}
with $X=S^{1}$.  But this follows immediately from the previous discussion,
especially~\thetag{B.48}.
        \enddemo

\newpage
%&amstex2.1
%**start of header
%&amstex2.1
%\documentstyle{amsppt}
%\input amstex.mac
%\NoRunningHeads
%\input screensize
\widestnumber\key{SSSSSSSSSSSSSS}   % for widest bibliography name
%**end of header


\Refs\tenpoint

\ref
\key AMM     
\by A. Alekseev, A. Mironov, A. Morozov
\paper On B-independence of RR charges
\finalinfo{\tt hep-th/0005244}
\endref

\ref 
\key A 
\by M. F. Atiyah
\paper Power operations in $K$-theory
\jour Quart. J. Math. 
\vol 17 
\yr 1966 
\pages 165--193 
\endref


\ref
\key AH      
\by M. F. Atiyah, F. Hirzebruch \paper Vector bundles and homogeneous spaces\jour Proc. Symp. Pure Math. \vol 3 \yr 1961 \pages 7--38 
\endref

\ref
\key BDS     
\by C. Bachas, M. Douglas, C. Schweigert 
\paper Flux stabilization of D-branes 
\jour JHEP 0005:048, 2000
\finalinfo {\tt hep-th/0003037}
\endref

\ref
\key BSS     
\by C. Bizdadea, L. Saliu, S. O. Saliu 
\paper On Chapline-Manton couplings: a cohomological approach 
\jour Phys. Scripta  
\vol 61 
\yr 2000 
\pages 307--310
\finalinfo {\tt hep-th/0008022}
\endref

\ref
\key B       
\by J. M. Bismut \paper The Atiyah-Singer Index Theorem for
families of Dirac operators: two heat equation proofs \jour Invent. math. \vol
83\yr 1986 \pages 91--151
\endref

\ref
\key BF      
\by J. M. Bismut, D. S. Freed \paper The analysis of elliptic
families I: Metrics and connections on determinant bundles \jour Commun. Math.
Phys. \vol 106 \pages 159--176 \yr 1986 
\moreref 
\paper The analysis of elliptic families II:
Dirac operators, eta invariants, and the holonomy theorem of Witten \jour
Commun. Math. Phys. \vol 107 \yr 1986 \pages 103--163
\endref

\ref 
\key BS 
\by A. Borel, J.-P. Serre 
\paper Le th\'eor\`eme de Riemann-Roch
\jour Bull. Soc. Math. France  
\vol 86 
\yr 1958 
\pages 97--136
\endref

\ref
\key Br      
\by W. Browder 
\paper The Kervaire invariant of framed manifolds and its genearlization 
\jour Annals of Math 
\vol 90 
\yr 1969 
\pages 157--186
\endref

\ref
\key Bry     
\by J.-L. Brylinski
\book Loop Spaces, Characteristic Classes and Geometric Quantization
\publ Birkh\"auser
\publaddr Boston
\yr 1993
\endref

\ref
\key CMW     
\by A.L. Carey, M.K. Murray, B.L. Wang
\paper Higher bundle gerbes and cohomology classes in gauge theories
\jour J.Geom.Phys.  
\vol 21 
\yr  1997 
\pages 183--197 
\finalinfo {\tt hep-th/9511169}
\endref

\ref
\key CM      
\by{Chapline, G. F. and Manton, N. S.}
\paper{Unification of {Y}ang-{M}ills theory and supergravity in
             ten dimensions}
\jour{Phys. Lett. B}
\vol{120}
\yr{1983}
\pages{105--109}
\endref

\ref
\key CS      
\paper Differential characters and geometric invariants
\by J. Cheeger, J. Simons 
\inbook Geometry and topology (College Park, Md., 1983/84) 
\pages 50--80 
\bookinfo Lecture Notes in Mathematics 
\vol 1167 
\publ Springer
\publaddr Berlin 
\yr 1985
\endref

\ref
\key CY      
\by Y.-K. E. Cheung and Z. Yin
\paper Anomalies, Branes, and Currents
\jour Nucl. Phys. \vol B517 \yr 1998 \pages 185-196
\finalinfo {\tt hep-th/9803931}
\endref

\ref
\key C       
\by S. Coleman 
\book Aspects of Symmetry 
\publ Cambridge University Press 
\yr 1985
\endref

\ref
\key D       
\by P. Deligne 
\paper Th\'eorie de Hodge. II 
\jour Inst. Hautes Etudes Sci. Publ. Math. 
\yr 1971 
\issue 40
\pages 5--57
\endref

\ref
\key DEFJKMMW
\book{Quantum Fields and Strings: A Course for Mathematicians}
\eds{P. Deligne, P. Etingof, D. S. Freed, L. C. Jeffrey, D. Kazhdan,
J. W. Morgan, D. R. Morrison, E. Witten}
 \publ{American Mathematical Society}
 \yr{1999}
 \publaddr{Providence, RI}
 \bookinfo{Volume~1}
\endref

\ref
\key DF      
\by{P. Deligne, D. S. Freed}
\paper{Classical field theory}
\inbook{Quantum Fields and Strings: A Course for Mathematicians}
 \eds{P. Deligne, P. Etingof, D. S. Freed, L. C. Jeffrey, D. Kazhdan,
J. W. Morgan, D. R. Morrison, E. Witten}
 \publ{American Mathematical Society}
 \yr{1999}
 \publaddr{Providence, RI}
 \bookinfo{Volume~1}
\pages{137--225}
\endref

\ref
\key F1      
\by D. S. Freed 
\paper Classical Chern-Simons theory
\jour Adv. Math.  
\vol 113 
\yr 1995 
\pages 237--303
\finalinfo {\tt hep-th/9206021}
\endref

\ref
\key F2      
\by D. S. Freed \paper On determinant line bundles \inbook Mathematical Aspects of String Theory \bookinfo ed. S. T. Yau \publ World Scientific Publishing \yr 1987 
\endref

\ref
\key FH      
\by D. S. Freed, M. J. Hopkins 
\paper On Ramond-Ramond fields and $K$-theory  
\jour J. High Energy Phys.  
\yr 2000 
\paperinfo Paper 44
\finalinfo{\tt hep-th/0002027}
\endref

\ref
\key FHMM    
\by D. S. Freed, J. A. Harvey, R. Minasian, G. Moore
\paper{Gravitational anomaly cancellation for {M}-theory fivebranes}
\jour{Adv. Theor. Math. Phys.}
\vol{2}
\yr{1998}
\pages{601--618}
\finalinfo {\tt hep-th/9803205}
\endref

\ref
\key FW      
\by D. S. Freed, E. Witten 
\paper Anomalies in string theory with $D$-branes
\jour Asian J. Math
\toappear 
\finalinfo{\tt hep-th/9907189}
\endref

\ref 
\key G 
\by P. Gajer 
\paper Geometry of Deligne cohomology
\jour Invent. Math.  
\vol 127 
\yr 1997 
\pages 155--207
\endref

\ref
\key GS      
\by M. B. Green, J. H. Schwarz 
\paper Anomaly cancellations in supersymmetric $D=10$ gauge theory and
superstring theory 
\jour Phys. Lett. B 
\yr 1984
\vol 149 
\pages 117--122
\endref

\ref
\key GSW     
\by M. B. Green, J. H. Schwarz, E. Witten 
\book Superstring theory, Volume 2 
\publ Cambridge University Press 
\yr 1987
\endref

\ref
\key H       
\by N. J. Hitchin 
\paper Lectures on special lagrangian submanifolds 
\finalinfo {\tt math.DG/9907034}
\endref

\ref
\key HS      
\by M. J. Hopkins, I. M. Singer 
\paper Quadratic functions in geometry, topology, and M-theory 
\toappear
\endref

\ref
\key K       
\by{D. Kazhdan}
\paper{Introduction to QFT}
\paperinfo{notes by Roman Bezrukavnikov}
\inbook{Quantum Fields and Strings: A Course for Mathematicians}
 \eds{P. Deligne, P. Etingof, D. S. Freed, L. C. Jeffrey, D. Kazhdan,
J. W. Morgan, D. R. Morrison, E. Witten}
 \publ{American Mathematical Society}
 \yr{1999}
 \publaddr{Providence, RI}
 \bookinfo{Volume~1}
\pages{377--418}
\endref

\ref 
\key L 
\by Lott, J.
\paper{$\RZ$ index theory}
\jour{Comm. Anal. Geom.}
\vol{2}
\yr{1994}
\pages{279--311}
\endref

\ref
\key MM      
\by R. Minasian, G. Moore 
\paper $K$-theory and Ramond-Ramond charge
\jour J. High Energy Phys.
\yr 1998
\finalinfo no.~11, Paper 2, 7 pp, {\tt hep-th/9710230}
\endref

\ref
\key MW      
\by G. Moore, E. Witten 
\paper Self-duality, Ramond-Ramond fields, and $K$-theory 
\jour J. High Energy Phys.
\finalinfo 0005 (2000) 032, {\tt hep-th/9912279}
\endref

\ref
\key Q       
\by D. Quillen \paper Superconnections and the Chern character \jour Topology
\vol 24 \yr 1985 \pages 89--95 
\endref

\ref
\key R       
\by L. H. Ryder
\book Quantum Field Theory 
\publ Cambridge University Press 
\yr 1985
\endref

\ref
\key T       
\by W. Taylor 
\paper D2-branes in $B$ fields
\jour JHEP 
\finalinfo 0007 (2000) 039, {\tt hep-th/0004141}
\endref

\ref
\key Wa      
\by F. W. Warner 
\book Foundations of Differentiable Manifolds and Lie Groups 
\publ Springer-Verlag 
\publaddr New York 
\yr 1983
\endref

\ref
\key W1      
\by E. Witten
\paper Overview of $K$-theory applied to strings
\finalinfo {\tt hep-th/0007175}
\endref

\ref
\key W2      
\by E. Witten 
\paper Five-brane effective action in M-theory 
\jour J. Geom. Phys. 
\vol 22 
\yr 1997 
\pages 103--133 
\finalinfo {\tt hep-th/9610234}
\endref

\ref
\key W3      
\by E. Witten 
\paper D-branes and $K$-theory 
\jour JHEP 
\vol 812:019 
\yr 1998 
\finalinfo {\tt hep-th/9810188}
\endref

\ref
\key W4      
\by E. Witten 
\paper Duality realtions among topological effects in string theory 
\jour JHEP
\finalinfo  0005 (2000) 031, {\tt hep-th/9912086}
\endref

\ref
\key W5      
\by E. Witten
\paper Baryons and branes in anti de Sitter space
\jour JHEP 
\finalinfo 9807 (1998) 006, {\tt hep-th/9805112}
\endref

\endRefs
\enddocument